\documentclass[onecolumn,twoside,superscriptaddress,nofootinbib,preprintnumbers,11pt,floatfix]{revtex4-2}
\usepackage{graphicx}
\usepackage{xcolor}
\usepackage{amsmath,amsfonts,amssymb}
\usepackage{physics}
\usepackage{hyperref}
\usepackage[normalem]{ulem}
\hypersetup{
    colorlinks=true,
    linkcolor=red,
    citecolor=blue,
    urlcolor=magenta
}
\usepackage{orcidlink}

\begin{document}

\title{Jet quenching and its substructure dependence due to color decoherence}

\author{Xiang-Pan Duan \orcidlink{0009-0001-0471-8832}}
\email{xpduan20@fudan.edu.cn}
\affiliation{Key Laboratory of Nuclear Physics and Ion-beam Application (MOE), Institute of Modern Physics, Fudan University, Shanghai 200433, China}
\affiliation{Shanghai Research Center for Theoretical Nuclear Physics, NSFC and Fudan University, Shanghai 200438, China}
\affiliation{Instituto Galego de F\'isica de Altas Enerx\'ias IGFAE, Universidade de Santiago de Compostela, E-15782 Galicia, Spain}

\author{Lin Chen \orcidlink{0000-0003-2082-3533}}
\email{lin.chen@usc.es}
\affiliation{Instituto Galego de F\'isica de Altas Enerx\'ias IGFAE, Universidade de Santiago de Compostela, E-15782 Galicia, Spain}

\author{Guo-Liang Ma \orcidlink{0000-0002-7002-8442}}
\email{glma@fudan.edu.cn}
\affiliation{Key Laboratory of Nuclear Physics and Ion-beam Application (MOE), Institute of Modern Physics, Fudan University, Shanghai 200433, China}
\affiliation{Shanghai Research Center for Theoretical Nuclear Physics, NSFC and Fudan University, Shanghai 200438, China}

\author{Carlos A. Salgado \orcidlink{0000-0003-4586-2758}}
\email{carlos.salgado@usc.es}
\affiliation{Instituto Galego de F\'isica de Altas Enerx\'ias IGFAE, Universidade de Santiago de Compostela, E-15782 Galicia, Spain}
\affiliation{Axencia Galega de Innovación (GAIN), Xunta de Galicia, Galicia, Spain}

\author{Bin Wu \orcidlink{0000-0002-3320-0442}}
\email{b.wu@cern.ch}
\affiliation{Instituto Galego de F\'isica de Altas Enerx\'ias IGFAE, Universidade de Santiago de Compostela, E-15782 Galicia, Spain}

\begin{abstract}
Motivated by color coherence and decoherence effects in the QCD medium, we propose a theoretical framework that combines vacuum-like emissions and medium-induced radiation to study jet quenching and its dependence on jet cone sizes and substructure. In our approach, a jet produced at a hard scale $Q$ first undergoes vacuum-like evolution, as described by the well-established generating-function method in the double logarithmic approximation. These vacuum-like emissions generate subjets at an infrared momentum scale $Q_0$. Each subjet then experiences medium-induced energy loss as described by the BDMPS-Z formalism. By modeling the QCD bulk medium using OSU (2+1)-dimensional viscous hydrodynamics and treating $Q_0$ together with the jet-quenching parameters at the initial proper time of the hydrodynamic evolution as free parameters, our approach provides a very good description of the inclusive jet modification factor $R_{AA}$ for large-radius jets and its dependence on jet substructure in 0-10\% PbPb collisions at $\sqrt{s_{NN}} = 5.02~\rm{TeV}$, as measured by the ATLAS experiment.
\end{abstract}

\maketitle

\newpage
\tableofcontents

\section{Introduction}
\label{sec:intro}

High-energy heavy-ion collisions at the Relativistic Heavy Ion Collider (RHIC)~\cite{STAR:2003pjh,BRAHMS:2004adc,PHOBOS:2004zne,PHENIX:2004vcz,STAR:2005gfr,Chen:2024aom} and the Large Hadron Collider (LHC)~\cite{ATLAS:2010isq,ALICE:2010yje,CMS:2011iwn,Shou:2024uga} create hot and dense quantum chromodynamics (QCD) matter known as the quark-gluon plasma (QGP)~\cite{Shuryak:1978ij,Gyulassy:2004zy}. A primary objective of heavy-ion collisions is to characterize the transport properties of such QCD matter. Partons produced in the initial hard scattering, with high transverse momentum ($p_T$) and large virtuality ($Q^2$), propagate through the QGP and undergo multiple interactions with medium partons. These interactions lead to parton energy loss and transverse momentum broadening, causing suppression of jet production compared with $pp$ collisions. This phenomenon is commonly referred to as jet quenching~\cite{Gyulassy:1990ye,Wang:1992qdg,Baier:1996sk,Salgado:2002cd,Armesto:2005iq,Casalderrey-Solana:2007knd,Qin:2015srf,Cao:2020wlm,Duan:2023gmp} and has become a key tool for probing the QGP properties. Measurements of jet observables such as the nuclear modification factor~\cite{Eskola:2004cr,Caucal:2020xad,Zhang:2021sua,JETSCAPE:2022jer,Luo:2023nsi,JETSCAPE:2024cqe,Xie:2024xbn,Han:2025ukx,Chen:2026gka,Dang:2026ezw}, dijet momentum imbalance~\cite{Qin:2010mn,Ma:2013pha,Milhano:2015mng,Chen:2016cof,Chen:2018fqu,Li:2024uzk}, and jet substructure~\cite{Chang:2017gkt,CMS:2017qlm,CMS:2018fof,ALICE:2021mqf,JETSCAPE:2023hqn,ALICE:2024jtb,Duan:2025wsy} provide valuable insight into the mechanisms of parton energy loss and the characteristics of QGP.

Among these observables, the nuclear modification factor $R_{AA}$ plays an essential role in jet quenching studies, as it provides a quantitative measurement of medium-induced modifications. The significant suppression of high-$p_T$ hadrons observed in the PHENIX~\cite{PHENIX:2001hpc} and STAR~\cite{STAR:2002ggv} experiments provided early evidence for jet quenching. Subsequent measurements at RHIC~\cite{STAR:2003fka,PHENIX:2005yls,PHENIX:2012jha,STAR:2020xiv} and the LHC~\cite{CMS:2012aa,CMS:2016uxf,CMS:2016xef,ATLAS:2018gwx,ALICE:2019qyj,CMS:2021vui,ATLAS:2023hso,ALICE:2023waz} further confirmed the suppression of high-$p_{T}$ hadrons and jets. With increasing collision energy and improved experimental precision, studies of $R_{AA}$ for inclusive jets have been extended to progressively higher jet transverse momentum, providing more stringent tests of theoretical descriptions of jet quenching. In particular, the ATLAS experiment~\cite{ATLAS:2018gwx,ATLAS:2023hso} has reported the $R_{AA}$ measurements for jet transverse momentum $p_T$ up to $1~\rm{TeV}$. For such observables, a key medium characteristic governing the jet quenching phenomenon is the jet transport coefficient $\hat{q}$~\cite{Baier:1996sk,Eskola:2004cr}, which quantifies the average transverse momentum squared (with respect to the jet direction) transferred from the medium to a propagating parton per unit path length. The $\hat{q}$ has been extensively investigated through theoretical calculations~\cite{Liu:2006ug,Armesto:2009zi,Chen:2010te,Majumder:2012sh,Panero:2013pla,JET:2013cls,Liu:2015vna}.

As the evolution parameter for both vacuum and medium, virtuality determines the resolvable parton multiplicity by parton shower process. Hence, a quantitative understanding of jet quenching requires a detailed description of jet evolution~\cite{Salgado:2003gb,Armesto:2003jh,Casalderrey-Solana:2011ule,Armesto:2011ir,Mehtar-Tani:2011lic,Mehtar-Tani:2012mfa,Blaizot:2013vha,Caucal:2018dla,Caucal:2019uvr,Caucal:2020uic,Adhya:2021kws}, including vacuum-like emissions and medium-induced radiation. In vacuum, jet evolution proceeds through parton shower process with its virtuality gradually decreasing, while parton multiplicity can be calculated according to the Dokshitzer–Gribov–Lipatov–Altarelli–Parisi (DGLAP) evolution equation~\cite{Gribov:1972ri,Dokshitzer:1977sg,Altarelli:1977zs}. However, in medium, jet evolution proceeds through both vacuum-like emissions and medium-induced radiation. Currently, jet energy loss in QCD medium can be calculated by different theory frameworks. The Baier–Dokshitzer–Mueller–Peigne–Schiff and Zakharov (BDMPS-Z) formalism~\cite{Baier:1996sk,Baier:1996kr,Zakharov:1996fv,Baier:1998kq} in the weak-coupling regime provides a comprehensive description of the medium-induced gluon radiation spectrum of high-energy partons due to the Landau–Pomeranchuk–Migdal (LPM) effects in a dense QCD medium. Within this formalism, multiple soft gluon emissions from high-energy partons are encoded in the quenching weights~\cite{Salgado:2002cd,Salgado:2003gb}, which describe the probability distribution for radiating an energy fraction that is approximately independent of the initial energy of the hard parton.
Due to color decoherence~\cite{Mehtar-Tani:2011hma} in the BDMPS-Z formalism, resolved partons lead to radiation as individual color charges. Parton multiplicity generated during hard splitting connects jet evolution to the dynamics of vacuum-like emissions and medium-induced radiation. Although multiplicity distribution has been investigated in several theoretical frameworks~\cite{Duan:2025ngi,Dokshitzer:2025owq,Dokshitzer:2025fky,Duan:2025lvi,Duan:2025gpp}, its relationship with virtuality and color decoherence in QGP has not been fully explored. Therefore, a deep understanding of this connection is crucial for understanding jet quenching effects.

Various alternative theoretical frameworks have been developed to describe jet quenching. For example, soft collinear effective theory with Glauber gluon interactions (SCET${_\text{G}}$)~\cite{Ovanesyan:2011xy,Kang:2014xsa} provides a systematic effective-field-theory framework for computing medium-modified jet observables. Higher-Twist (HT) approach~\cite{Guo:2000nz,Zhang:2003yn} compute medium-induced radiative energy loss through twist expansions. Furthermore, other frameworks, such as JEWEL~\cite{Zapp:2008gi,Zapp:2013vla}, MARTINI~\cite{Schenke:2009gb}, Hybrid~\cite{Casalderrey-Solana:2014bpa}, LBT~\cite{He:2015pra}, JETSCAPE~\cite{JETSCAPE:2017eso}, and LIDO~\cite{Ke:2018jem}, have been widely used for the phenomenological analysis of RHIC and LHC data. In these frameworks, vacuum-like emissions and medium-induced radiation are incorporated in different ways, reflecting the formidable challenges of performing a complete, first-principles calculation of jet evolution in a QCD medium.

In this work, we aim to establish a connection between virtuality evolution and the suppression of inclusive jets in high-energy heavy-ion collisions, exemplified through phenomenological studies of $R_{AA}$ for large-radius jets and its dependence on jet substructure. A consistent treatment of multiplicity, virtuality, and decoherence is particularly important for explaining the jet $R_{AA}$ at high $p_T$. Although the $R_{AA}$ distributions have been studied by various energy loss models~\cite{JET:2013cls,Zapp:2012ak,Chien:2015hda,He:2018xjv,Andres:2019eus,Adhya:2019qse,Ke:2020clc,Caucal:2020uic,Mehtar-Tani:2021fud,JETSCAPE:2021ehl,Mehtar-Tani:2024jtd,Kudinoor:2025ilx,Han:2025ukx,Datta:2025gql}, the underlying physical assumptions differ significantly. These differences lead to discrepancies in the predicted jet results. Hence, establishing the relationship between multiplicity, virtuality evolution, and color decoherence is essential for reducing model discrepancies.

To this end, we build a framework that incorporates color decoherence effects in jet-medium interactions by identifying the number of independent subjets that suffer energy loss during the virtuality evolution of vacuum-like emissions. These subjest are associated with parton multiplicity at the color decoherence scale, calculated using the double logarithmic approximation (DLA)~\cite{Duan:2025ngi,Duan:2025lvi}. Radiative energy loss is computed within the BDMPS-Z formalism~\cite{Baier:1996sk,Baier:1996kr,Zakharov:1996fv,Baier:1998kq}, while the QGP evolution is modeled using the OSU (2+1)-dimensional viscous hydrodynamic model~\cite{Song:2007ux,Song:2008si,Song:2010mg}. Based on this framework, we perform the $R_{AA}$ calculation of inclusive jets in 0-10\% PbPb collisions at $\sqrt{s_{NN}} = 5.02~\rm{TeV}$, comparing our results with ATLAS measurements~\cite{ATLAS:2023hso}. In particular, we focus on the large-radius jet suppression and its dependence on substructure. Furthermore, our study links the internal structure of jet evolution to jet quenching effects. By elucidating the roles of virtuality evolution and color decoherence in jet energy loss, we aim to improve the theoretical explanation of QGP properties and provide new pathways for medium-induced energy loss.

The paper is organized as follows. Section~\ref{sec2} presents our theoretical approach, incorporating parton multiplicity within the DLA, radiative energy loss within the BDMPS-Z formalism, the leading-order (LO) differential jet cross section, motivated by color decoherence effects. Section~\ref{sec3} presents the numerical results for jet energy loss. Our main results for $R_{AA}$ of large-radius jets reclustered from smaller-radius subjets, including comparisons with the ATLAS data in ref.~\cite{ATLAS:2023hso} for the large-radius dependence of jet substructure suppression as well as the behavior of single versus multiple subjets, are presented in Sections~\ref{sec4} and \ref{sec5}, respectively. Finally, we summarize our findings and discuss future directions in Section~\ref{sec6}.

\section{Jet suppression with color decoherence}
\label{sec2}

In this work, motivated by color coherence and decoherence~\cite{Mehtar-Tani:2011hma,Casalderrey-Solana:2011ule}, we propose a theoretical approach to describe jet quenching in AA collisions by incorporating both vacuum-like emissions and medium-induced radiation: the parton shower first evolves from high to low virtuality as vacuum-like emissions, after which the resulting parton multiplicity enhances jet energy loss in QCD medium.

\subsection{Building blocks: vacuum-like emissions and medium-induced radiation}

For completeness, we first provide a concise summary of both vacuum-like emissions and medium-induced radiation, which will be used in our phenomenological analyses. For vacuum-like emissions, we use the generating functions to present the parton shower process within the DLA~\cite{Dokshitzer:1991wu}, which has been recently employed and extended to investigate KNO scaling within QCD jets in $pp$ collisions~\cite{Duan:2025ngi,Dokshitzer:2025owq, Duan:2025lvi}. For medium-induced radiation, we employ the jet energy loss formula using the quenching weights~\cite{Baier:2001yt, Salgado:2003gb} within the BDMPS-Z formalism~\cite{Baier:1996sk,Baier:1996kr,Zakharov:1996fv, Baier:1998kq}, which provides a systematic description of multiple soft scatterings in the dense QGP matter, as has been tested with experimental data.

\subsubsection{The virtuality evolution of multiplicity distributions in jets}

The virtuality evolution of multiplicity distributions in jets can be formulated using generating functions. For a parton of flavor $i$, produced at an initial scale $Q$, the generating function for multiplicity distributions is defined as
\begin{align}
    Z_i(u,Q) \equiv \sum_{n=0}^{\infty} u^n P_i(n,Q)
\end{align}
where the multiplicity probability distribution $P_i(n, Q)$ denotes the probability of finding $n$ particles in the shower at the scale $Q$. In this work, we focus on parton multiplicity distributions evolving from $Q$ down to a lower scale $Q_0$.

Introducing the logarithmic evolution variable $y \equiv \ln(Q/Q_0)$, the generating functions in DLA can be recast into an integral form~\cite{Dokshitzer:1991wu}:
\begin{align}\label{eq:GF_DLA}
    Z_i(u, y) = u \exp \left\{ c_i \int_0^y d\bar{y}~(y - \bar{y}) \gamma_0^2 [Z_g(u,\bar{y})-1] \right\},
\end{align}
where $\bar{y} \equiv \ln(k_\perp/Q_0)$ denotes the logarithmic transverse-momentum scale of the emitted gluon, and $\gamma_0 = \sqrt{2N_c\alpha_s/\pi}$ is the anomalous dimension. The color factor ratio is $c_g = 1$ for gluon jets and $c_q \equiv C_F/C_A$ for quark jets, with the Casimir factors $C_A = N_c$ and $C_F = (N_c^2-1)/(2N_c)$. Note that in the DLA one considers only collinear and soft gluon radiation. Therefore, at the end of the parton shower all partons in a gluon jet are gluons, whereas in a quark jet all partons except the leading one are gluons because the jet is by definition initiated by a quark or an antiquark.

The multiplicity probability distribution is obtained from the generating functions through successive differentiation with respect to the auxiliary variable $u$,
\begin{align}\label{eq:Pndef}
    P_i(n,Q) = \frac{1}{n!} \frac{\partial^n}{\partial u^n} Z_i(u,Q) \bigg|_{u=0}.
\end{align}
From the generating functions given in eq.~\eqref{eq:GF_DLA}, one can further derive a set of recursive relations for the parton multiplicity distributions $P_i(n,Q)$~\cite{Duan:2025ngi,Duan:2025lvi}:
\begin{align}\label{eq:Pqg}
    P_i(1,Q)
    &= \exp \left\{-c_i \int_0^y d\bar{y}~(y-\bar{y}) \gamma_0^2 \right\}, \nonumber\\
    P_i(n+1,Q)
    &= c_i \sum_{k=1}^{n} \frac{k}{n}P_i(n+1-k,Q) \int_0^y d\bar{y}~(y-\bar{y}) \gamma_0^2 P_g(k,\bar{y}).
\end{align}
And these distributions satisfy the normalization condition
\begin{align}
\label{eq:Pn_normalization}
    \sum_{n=1}^{\infty} P_i(n,Q) = 1.
\end{align}

\subsubsection{Quenching weights within the BDMPS-Z formalism}

In this work, the medium-induced energy loss of a hard parton of flavor $i$ is taken to be that given by the BDMPS-Z formalism. Specifically, the parton energy loss is described by the probability distribution $D_i(\epsilon)$, which gives the probability that the parton loses energy $\epsilon$ due to multiple soft gluon radiation induced by multiple scatterings in the medium~\cite{Baier:2001yt, Salgado:2003gb}.

For a hard parton propagating through the medium, the medium-induced gluon radiation spectrum due to the LPM effect is given by~\cite{Baier:1998kq}
\begin{align}
    \omega \frac{dI_i}{d\omega}
    = \frac{2\alpha_s C_i}{\pi}K_i(x)\,
      \ln \left|\cos\!\left[(1+i)
      \bigg(\frac{\omega_{c}}{2\omega}
      \frac{1-x + C_i x^2/N_c}{1-x}\bigg)^{\frac{1}{2}}
      \right]\right|,
\end{align}
where $x$ denotes the energy fraction carried by the radiated gluon, $\omega$ denotes its energy, and the color factor
$C_i=C_F=4/3$ for quarks ($i=q$) and antiquarks ($i=\bar q$), and
$C_i=C_A=3$ for gluons ($i=g$). The splitting function is
\begin{align}
    K_i(x) =
    \begin{cases}
    \frac{1}{2}\big[1+(1-x)^2\big], & \text{for $i=q,\bar q$}, \\
    x\big[\frac{x}{1-x}+\frac{1-x}{x}+x(1-x)\big],
    & \text{for $i=g$}.
    \end{cases}
\end{align}
Here, the splitting function is specific to the medium-induced radiation formalism and does not affect the vacuum-like emissions treated in the DLA. As discussed in the next subsection, the vacuum-like evolution occurs before medium-induced radiation becomes dominant.

Given a static medium of finite length $L$, the characteristic gluon frequency is
\begin{align}\label{eq:omegac}
    \omega_{c} \equiv \frac{1}{2}\, \hat{q}\, L^2,
\end{align}
which sets the typical energy scale of medium-induced radiation. Here and throughout the following discussions, $\hat{q}$ denotes the jet transport coefficient for a gluon. Generically, for a parton of flavor $i$, the jet transport coefficient $\hat{q}_i$, which governs both transverse momentum broadening and the induced radiation spectrum, is defined as~\cite{Baier:1996sk}
\begin{align}\label{eq:qhat}
    \hat{q}_i
    = \rho \int dq_T^2\, q_T^2\, \frac{d\sigma_i}{dq_T^2}= \frac{1}{\lambda_i}\int dq_T^2\, q_T^2\, \frac{1}{\sigma_i}\frac{d\sigma_i}{dq_T^2},
\end{align}
where $\rho$ is the medium density, $d\sigma_i/dq_T^2$ is the differential cross section for scattering between the high-energy parton and a medium constituent, and $\lambda_i\equiv 1/(\rho\sigma_i)$ denotes the mean free path of the parton.

\begin{figure}[htbp]
    \centering
    \includegraphics[width=0.6\textwidth]{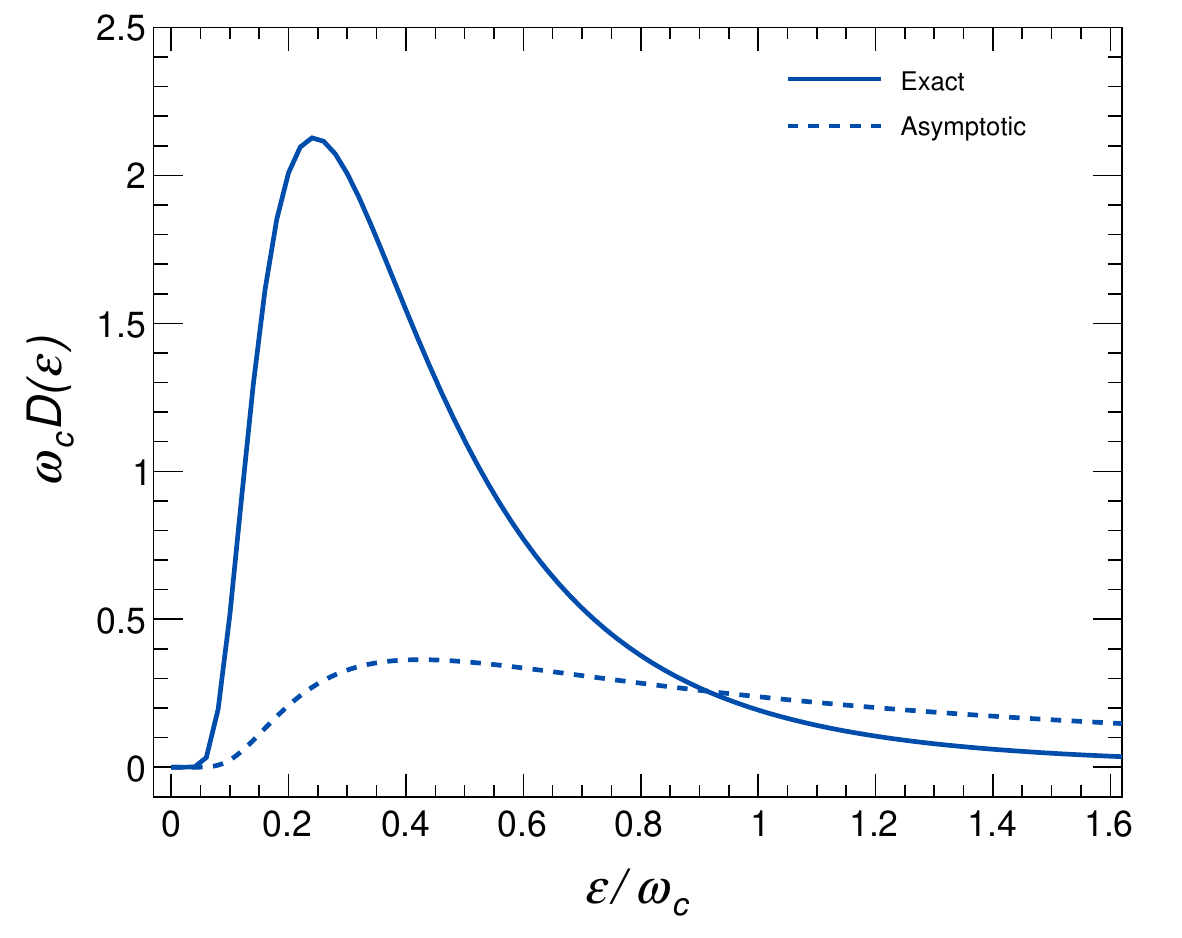}
    \caption{The probability distribution for the total medium-induced energy loss. Here, we show the numerical results of $\omega_c D(\epsilon)$ as a function of $\epsilon/\omega_c$ for $\alpha_s = 1/3$, along with the asymptotic solution in eq.~\eqref{eq:Dasy} from ref.~\cite{Baier:2001yt}.}
    \label{fig:Dcmp}
\end{figure}

Assuming independent soft gluon emissions (with $x\to 0$), the probability distribution for the total medium-induced energy loss $\epsilon$ of the high-energy parton is
\begin{align}\label{eq:D}
    D_i(\epsilon)
    = \sum_{n=0}^{\infty} \frac{1}{n!}
      \left[\prod_{k=1}^{n} \int d\omega_k \frac{dI_i(\omega_k)}{d\omega} \right]
      \delta\!\left(\epsilon - \sum_{k=1}^{n} \omega_k\right)
      \exp\!\left[- \int_0^\infty d\omega\, \frac{dI_i(\omega)}{d\omega} \right].
\end{align}
$D_i(\epsilon)$ can only be solved numerically~\cite{Salgado:2003gb}, with an asymptotic form given in ref.~\cite{Baier:2001yt}:
\begin{align}\label{eq:Dasy}
    \epsilon D_i(\epsilon)
    = \alpha_i \sqrt{\frac{\omega_c}{2\epsilon}} \exp\bigg\{-\frac{\pi\alpha_i^2 \omega_c}{\epsilon}\bigg\}
    \qquad
    \text{with }\alpha_i \equiv \frac{2\alpha_s C_i}{\pi}.
\end{align}
This approximation shows that the most probable (typical) energy loss of a high-energy parton is of order $\alpha_s^2 \omega_c$ instead of the average energy loss $\sim \alpha_s \omega_c$. To obtain the exact result, we solve $D_i$ using a Fourier transform:
\begin{align}
    D_i(\epsilon) 
    &= \int_{-\infty}^{\infty} \frac{d\nu}{2\pi}\, \exp\left[i\nu\epsilon- \int_0^\infty d\omega\, \frac{dI_i(\omega)}{d\omega}\, \left(1-e^{-i\nu \omega}\right)\right]\notag\\
    & = \int_{0}^{\infty}\frac{d\nu}{\pi}\, \exp\left\{-\int_0^\infty d\omega\, \frac{dI_i(\omega)}{d\omega}\, \left[1-\cos(\nu \omega)\right]\right\}
    \, \cos\left(\nu\epsilon- \int_0^\infty d\omega\, \frac{dI_i(\omega)}{d\omega }\, \sin(\nu \omega)\right).
\end{align}
This expression is equivalent to the inverse Laplace representation used in ref.~\cite{Salgado:2003gb}. By exploiting the analyticity of the Laplace-space function, the integration contour can be deformed to the imaginary axis, yielding the Fourier representation adopted here. Figure~\ref{fig:Dcmp} demonstrates a sizable quantitative deviation from the asymptotic form in eq.~\eqref{eq:Dasy}.

\subsection{Jet quenching incorporating vacuum-like emissions and medium-induced radiation}

In this section, we incorporate both vacuum-like emissions and medium-induced radiation, summarized above, into the jet evolution in heavy-ion collisions as follows: a high-$p_T$ parton produced in a hard scattering undergoes a parton shower in which its virtuality evolves from an initial jet scale $Q = p_T R$ down to a lower value $Q_0$, where $p_T$ is the initial jet transverse momentum and $R$ is the jet radius. In vacuum, $Q_0$ can be understood as roughly the hadronization scale. In the presence of a QCD medium, we take $Q_0$ as the typical scale below which the medium does not resolve further parton virtuality evolution; that is, the scale separating color coherence and decoherence effects.

Within this evolution, the multiplicity generated during the parton shower plays a crucial role in determining the magnitude of the energy loss, since a larger number of partonic constituents leads to increased interactions with the medium. When these partonic branches become resolved by the medium and lose their mutual color coherence, the corresponding enhancement of energy loss can be interpreted as a manifestation of color decoherence.

\subsubsection{Initial states of jet cross sections in heavy-ion collisions}

A quantitative characterization of jet quenching in AA collisions requires a pQCD baseline from $pp$ collisions. At LO, inclusive jet production arises from $2\to2$ partonic scatterings: $ab \to cd$, whose cross sections factorize into parton distribution functions (PDFs) and hard-scattering matrix elements $|\mathcal{M}_{ab\to cd}|^2$. Here, $ab$ and $cd$ denote the two incoming and outgoing partons, respectively. Medium-induced energy loss modifies the correspondence between the initial hard parton energy and the observed jet transverse momentum. In the following, we compute the nuclear modification factor $R_{AA}$ by convoluting the $pp$ jet spectra with the energy-loss probability distributions $D_i(\epsilon)$ of the parton flavor dependence, thereby naturally incorporating color decoherence effects discussed in the previous section.

The inclusive jet spectrum in $pp$ collisions at LO is given by
\begin{align}\label{eq:pp_sigma}
    \frac{d\sigma^{pp}_i}{dp_T}
    = 2p_T \sum_{a,b,c,d}(\delta_{ic} + \delta_{id}) \int dy_c\, dy_d\, x_a f_{a/p}(x_a,\mu^2)\, x_b f_{b/p}(x_b,\mu^2)\, \frac{d\hat{\sigma}_{ab\rightarrow cd}}{d\hat{t}},
\end{align}
where $a,b,c,d$ run over gluons ($g$) and the light quarks ($u,d,s$) and their antiquarks, $y_c$ and $y_d$ denote the rapidities of the outgoing partons, and the factorization and renormalization scales are chosen as $\mu^2 = p_T^2$. To obtain flavor-resolved jet spectra, each partonic channel $ab\to cd$ is classified according to whether the observed jet originates from a quark or a gluon, allowing the construction of $d\sigma_i^{pp}/dp_T$ with $i=q,\bar{q}$ or $g$. The longitudinal momentum fractions $x_a$ and $x_b$ of the incoming partons are determined by
\begin{align}
    x_a=\frac{p_T}{\sqrt{s}}\left(e^{y_c}+e^{y_d}\right), 
    \qquad
    x_b=\frac{p_T}{\sqrt{s}}\left(e^{-y_c}+e^{-y_d}\right).
\end{align}
The partonic cross sections $d\hat{\sigma}_{ab\rightarrow cd}/d\hat{t}$ depend on the Mandelstam variables,
\begin{align}
    \hat{s}=x_a x_b s, 
    \qquad
    \hat{t}=-x_a p_T \sqrt{s}\, e^{-y_c},
    \qquad
    \hat{u}=-x_b p_T \sqrt{s}\, e^{y_c},
\end{align}
with explicit expressions available in standard pQCD references~\cite{Peskin:1995ev}. In our numerical analysis, we employ the CT18NLO PDFs via \texttt{LHAPDF}~\cite{Buckley:2014ana} as the baseline for $pp$ collisions.

In AA collisions, the collinear PDFs in $pp$ collisions are replaced by transverse phase-space PDFs, whose Fourier transform is called the thickness beam function in ref.~\cite{Wu:2021ril}. Since the observables studied in this work are insensitive to the transverse momentum, we only require the transverse spatial distributions of partons, i.e., impact-parameter–dependent PDFs:
\begin{align}\label{eq:tpspdfs}
    f_{a/A}(\vec{r}, x)
    &=\mathcal{T}_{a/A}(\vec{r}, x, \vec{x})\bigg|_{\vec{x}=\vec{0}}\notag\\
    &\to T_A(\vec{r})\left[\frac{Z}{A}f_{a/p}(x,\mu^2)+\left(1-\frac{Z}{A}\right)f_{a/n}(x,\mu^2)\right]\qquad\text{in the Glauber model},
\end{align}
where $\mathcal{T}_{a/A}$ denotes the thickness beam function for finding parton $a$ in nucleus $A$, $x$ stands for the longitudinal momentum fraction carried by parton $a$, $\vec r$ denotes its transverse spatial position, and $\vec x$ is the coordinate conjugate to its transverse momentum. In the second line we show the reduced expression of the spatial PDFs as a product of the nuclear thickness function
\begin{align}
    T_A(\vec{r})
    \equiv\int dz \rho_A(\vec{r}, z),
\end{align}
where $\rho_A$ is the Woods–Saxon nuclear density distribution, and the averaged PDFs per nucleon constructed from the proton PDFs $f_{a/p}$ and neutron PDFs $f_{a/n}$ within the Glauber modeling of heavy nuclei~\cite{Miller:2007ri}, i.e., treating nuclei as collections of uncorrelated nucleons. The nucleus of mass number $A$ is assumed to contain $Z$ protons.

Note that even within the Glauber model, the parton distributions inside nuclei differ from the naive superposition of those of free protons and neutrons, since partons produced in the hard process of a binary nucleon–nucleon collision may subsequently encounter and interact with other nucleons~\cite{Armesto:2024rtl}. In this work, we adopt a phenomenological approach~\cite{Emelyanov:1999pkc,Hirano:2003pw} by incorporating cold nuclear effects through the impact-parameter–dependent nuclear modification factor $R_{a/A}(x,\mu^2,\vec{r})$:
\begin{align}\label{eq:Rp}
    R_{a/A}(x,\mu^2,\vec{r})
    = 1 + \left[R_{a/A}(x,\mu^2)-1\right] \frac{A\,T_A(\vec{r})}{\int d^2\vec{r}\,T_A^2(\vec{r})}.
\end{align}
The nuclear modification factor $R_{a/A}(x,\mu^2)$ is taken from the EPPS21 parametrization~\cite{Eskola:2021nhw}. The resulting spatially dependent nuclear PDFs can then be written as
\begin{align}\label{eq:NPDFs}
    f_{a/A}(\vec{r}, x) 
    = T_A(\vec{r}) \bar{f}_{a/A}(x,\mu^2,\vec{r}),
\end{align}
where CT18ANLO PDFs are used for both proton and neutron distributions. The average PDFs per nucleon at impact parameter $\vec{r}$ are defined as
\begin{align}
    \bar{f}_{a/A}(x,\mu^2,\vec{r})
    \equiv R_{a/A}(x,\mu^2,\vec{r}) \left[\frac{Z}{A}f_{a/p}(x,\mu^2)+\left(1-\frac{Z}{A}\right)f_{a/n}(x,\mu^2)\right].
\end{align}
It is worth noting that the coefficient of $T_A(\vec{r})$ in eq.~\eqref{eq:Rp} can in principle be determined directly from LHC data~\cite{Helenius:2012wd}.

\subsubsection{Jet cross sections in heavy-ion collisions}

Given the transverse phase-space PDFs expressed in terms of the nucleon PDFs in eq.~\eqref{eq:NPDFs}, we further assume that the initial-state and final-state effects can be factorized. Under this assumption, the jet cross section can be written in factorized form.

\begin{figure}[htbp]
    \centering
    \includegraphics[width=0.8\textwidth]{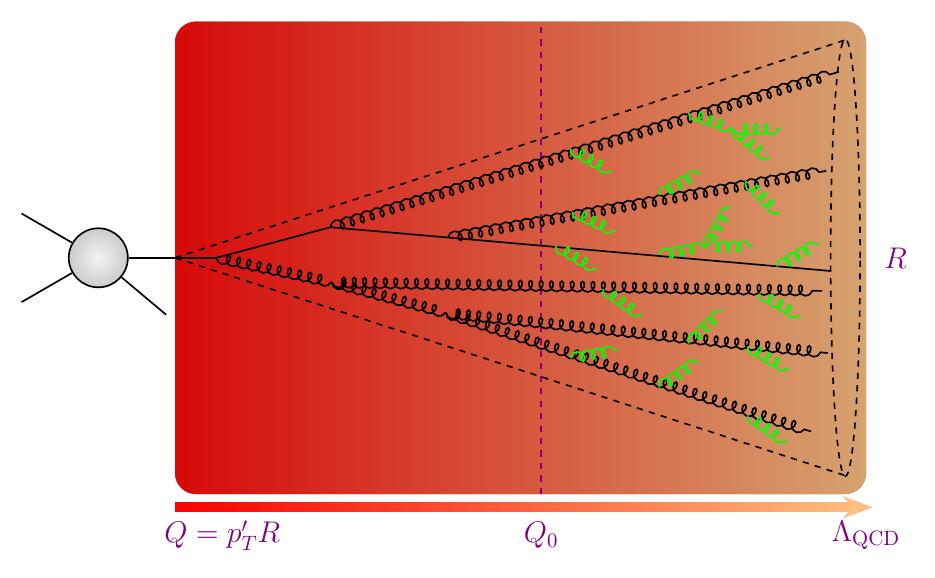}
    \caption{Jet evolution in the QCD medium. The black branching lines represent vacuum-like emissions from the initial virtuality $Q=p_T^{\prime}R$ down to the scale $Q_0$. Below $Q_0$, color decoherence allows the medium to resolve the jet substructure, and the resulting subjets propagate independently. The green lines represent soft gluons emitted through medium-induced radiation from the resolved subjets. The jet radius is indicated by $R$.}
    \label{fig:QCDmedium}
\end{figure}

\paragraph{Jet quenching with color decoherence.} In this work, as illustrated in figure~\ref{fig:QCDmedium}, we focus on the scenario in which the jet is resolved by the medium, with its constituents resolved down to a scale $Q_0$, treated as a parameter to be determined from experimental data in the following sections. Since we consider only high-$p_T$ jets, significant angular deviations of these subjets (partons of virtuality $Q_0$) from their vacuum-like emission vertices are not expected. Accordingly, following the approach used for jets in $pp$ collisions~\cite{Duan:2025ngi, Duan:2025lvi}, we choose $p_T^{\prime}R$ as the initial jet scale $Q$, where $R$ is the jet cone radius. This choice sets the scale for the vacuum-like evolution, which produces collinear radiation inside the jet cone and evolves from the hard scale down to the resolution scale $Q_0$. Here, $p_T^{\prime}$ denotes the initial transverse momentum of the parton produced in the hard process, defined as the sum of the final-state jet $p_T$ and the medium-induced energy loss.
\footnote{
Here, we employ the standard BDMPS-Z energy-loss formalism. Based on studies showing that medium-induced radiation can rapidly undergo multiple branchings into soft gluons that subsequently thermalize in the medium (see, e.g., \cite{Iancu:2015uja}), the radiated energy is treated as lost from the reconstructed jet. Consequently, the energy-loss formalism is independent of the jet cone size. A more systematic treatment incorporating $R$-dependent energy loss is beyond the scope of the present work and will be investigated in a future study.} 
In this framework, $Q_0$ plays a central role: it defines the threshold scale at which medium-induced energy loss becomes effective. A highly virtual parton produced at scale $Q$ undergoes successive splittings while propagating through the QGP medium, during which its virtuality decreases continuously through vacuum-like emissions. Once the virtuality reaches $Q_0$, partons begin to lose energy via multiple soft gluon radiation. Since virtualities above $Q_0$ are expected to be larger than any characteristic medium momentum scale, medium-induced modifications of the vacuum-like evolution are neglected. This transition therefore provides a natural separation between the vacuum-dominated regime and the medium-dominated regime, with energy loss considered only after partons of virtuality $Q_0$ have been produced.

According to the above physical picture, the differential cross section in AA collisions can be expressed as
\begin{align}\label{eq:sigmaAA_de}
    \frac{d\sigma^{AA}}{d^2 \vec{b}\, dp_T}
    &= \int d^2 \vec{r}\ d^2 \vec{b}\, T_A(\vec{r}+\vec{b}/2) T_B(\vec{r}-\vec{b}/2) \int \frac{d\phi}{2\pi} \nonumber\\
    &\quad \times \sum_{i} \Bigg[
        P_i(1,p_T^{\prime}R) 
        \int d\epsilon_1 D_i(\epsilon_1)
        \left.\frac{d\sigma_i^{NN}}{dp_T^{\prime}}\right|_{p_T^{\prime}=p_T+\epsilon_1} \nonumber\\
    &\quad + \sum_{n=2}^{N} P_i(n,p_T^{\prime}R)
        \int d\epsilon_1 D_i(\epsilon_1) 
        \left(\prod_{m=2}^{n} \int d\epsilon_m D_g(\epsilon_m)\right)
        \left.\frac{d\sigma_i^{NN}}{dp_T^{\prime}}\right|_{p_T^{\prime}=p_T+\sum_{k=1}^{n} \epsilon_k}
        \Bigg],
\end{align}
where $i=q,\bar{q}$ or $g$ denotes the type of the parton that initiates the jet, $\phi$ denotes the azimuthal angle of the jet transverse momentum, and the averaged inclusive jet cross section per nucleon pair is defined as
\begin{align}
    \frac{d\sigma_i^{NN}}{dp_T^{\prime}}
    = 2p_T^{\prime} \sum_{a,b,c,d} (\delta_{ic} + \delta_{id})\int dy_c dy_d x_i \bar{f}_{a/A}(x_i,\mu^2,\vec{r}+\vec{b}/2) x_b \bar{f}_{b/A}(x_b,\mu^2,\vec{r}-\vec{b}/2)\frac{d\hat{\sigma}_{ab \rightarrow cd}}{d\hat{t}}.
\end{align}
Here, the medium path length traversed by the jet depends on $\phi$, the azimuthal angle of the jet transverse momentum. Consequently, the energy-loss distribution $D_i$ also depends on $\phi$ through its dependence on $\omega_c$. Besides, the QCD medium is assumed to be longitudinally boost invariant. In this case, one can effectively consider the modification of $p_T$ due to radiative energy loss~\cite{Baier:1998yf,Iancu:2018trm}, which can be straightforwardly understood by boosting the jets to midrapidity at $\eta=0$.

Accordingly, the nuclear modification factor at impact parameter $\vec{b}$ is defined as
\begin{align}
    R_{AA}
    = \frac{1}{T_{AB}(\vec{b})}\, \frac{d\sigma^{AA}/d^2 \vec{b}\, dp_T}{d\sigma^{pp}/dp_T},
\end{align}
with the nuclear overlap function
\begin{align}
    T_{AB}(\vec{b})
    \equiv \int d^2 \vec{r}\, T_A(\vec{r}+\vec{b}/2)\, T_B(\vec{r}-\vec{b}/2).
\end{align}
In the following phenomenological studies, we run both the hydrodynamic simulation of the bulk QCD medium and the jet-quenching calculation at a single value of the impact parameter corresponding to the centrality class under study, with $|\vec{b}|$ taken as the average value reported in ref.~\cite{Loizides:2017ack}.

In phenomenological studies, we use the same nuclear PDFs from eq.~\eqref{eq:sigmaAA}, together with the CT18ANLO PDFs and the EPPS21 nPDF parametrizations. The calculations account for color decoherence, and the resulting theoretical predictions describe the experimental data well across the full $p_T$ range. In eq.~\eqref{eq:sigmaAA_de}, the second line corresponds to the energy-loss probability of a single parton, while the third line corresponds to the energy-loss probability of multiple partons. This approach yields the results presented in section~\ref{sec5}.

\paragraph{Jet quenching as a single color charge.} We now consider the simplest scenario in which jets cannot be resolved by the medium, so that jet energy loss can be treated as arising from the color-coherent propagation of a single color charge. In this scenario, the energy loss will be independent of the number of partons inside the jets and, therefore, the jet cross section will be independent of $P_i$ according to the normalization condition in eq.~\eqref{eq:Pn_normalization}. As a result, the inclusive jet cross sections differential in impact parameter $\vec{b}$ and jet transverse momentum $p_T$ in AA collisions are obtained by convoluting the $pp$ baseline with the energy-loss probability distribution $D_i(\epsilon)$ and integrating over the nuclear geometry,
\begin{align}\label{eq:sigmaAA}
    \frac{d\sigma^{AA}}{d^2 \vec{b}\, dp_T} 
    = \int d^2 \vec{r}\, T_A(\vec{r}+\vec{b}/2) T_B(\vec{r}-\vec{b}/2)\int \frac{d\phi}{2\pi}
    \sum_{i} \int d\epsilon D_i(\epsilon) \left.\frac{d\sigma_i^{NN}}{dp_T^{\prime}}\right|_{p_T^{\prime}=p_T+\epsilon},
\end{align}
as first studied in ref.~\cite{Baier:2001yt}. We take this as a baseline for comparison with the color-decoherent scenario.

\section{Jet energy loss with color decoherence}
\label{sec3}

In this section, we first present a detailed numerical analysis of parton multiplicity distributions for QCD jets. The calculations are performed within the DLA. Then, we calculate medium-induced energy loss with color decoherence effects in comparison with that for single color charges, which is calculated through quenching weights within the BDMPS-Z formalism~\cite{Baier:2001yt, Salgado:2003gb}.

\subsection{Multiplicity distributions in QCD jets}

The parton multiplicity distributions obtained from eq.~\eqref{eq:Pqg} have been investigated for $pp$ collisions in our previous works~\cite{Duan:2025ngi,Duan:2025lvi}. In particular, their scaling properties were analyzed in the context of KNO scaling for QCD jets. The parameter $Q_0$ is the infrared cutoff for the parton shower evolution. As detailed in the previous section, we extend this study to heavy-ion collisions in this work. We treat the early stage of jet evolution as vacuum-like emissions. These emissions dominate the dynamics before the jet interacts strongly with the medium down to the scale $Q_0$.

\begin{figure}[htbp]
    \centering
    \includegraphics[width=0.49\textwidth]{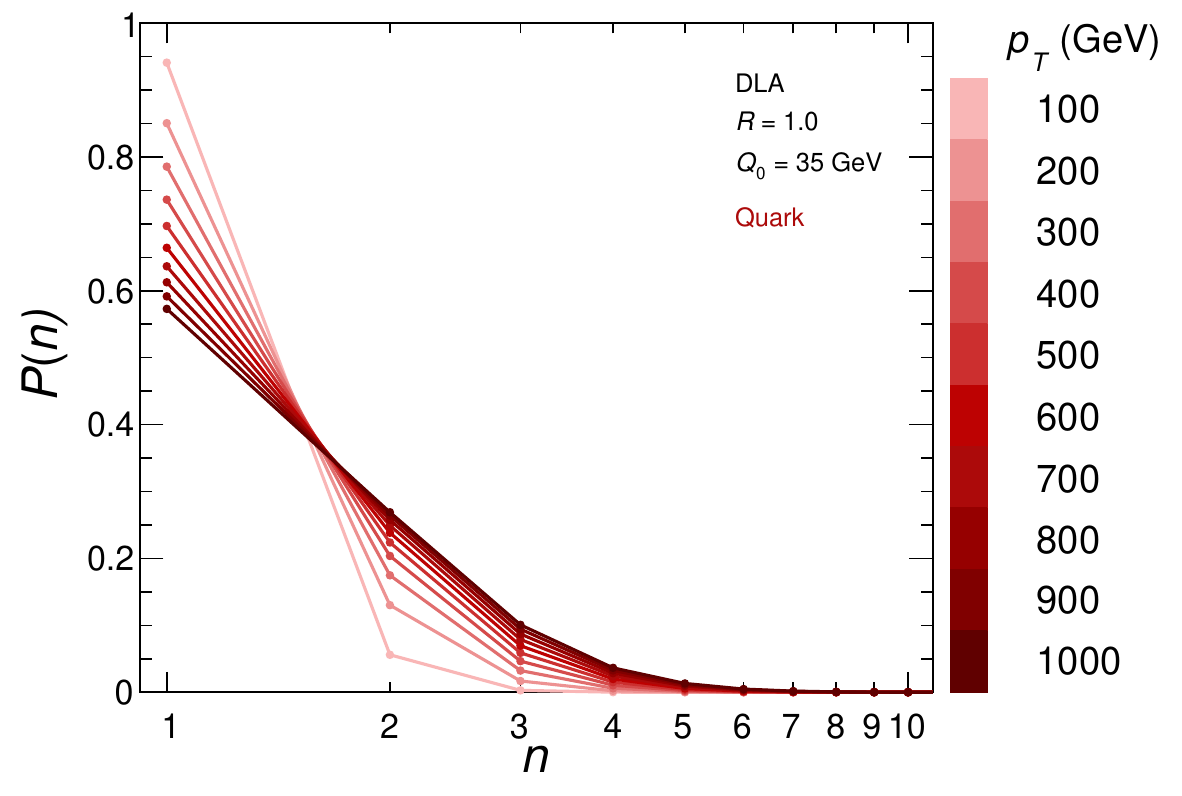}
    \includegraphics[width=0.49\textwidth]{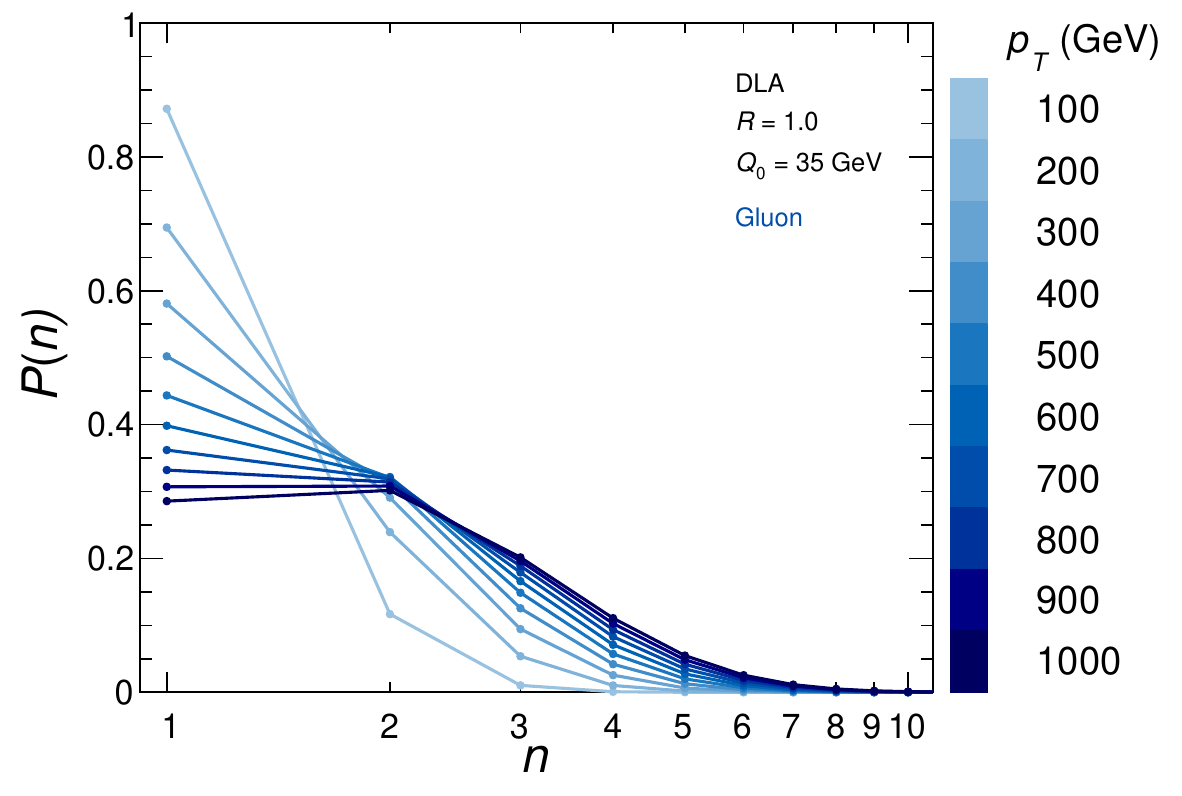}\\
    \includegraphics[width=0.49\textwidth]{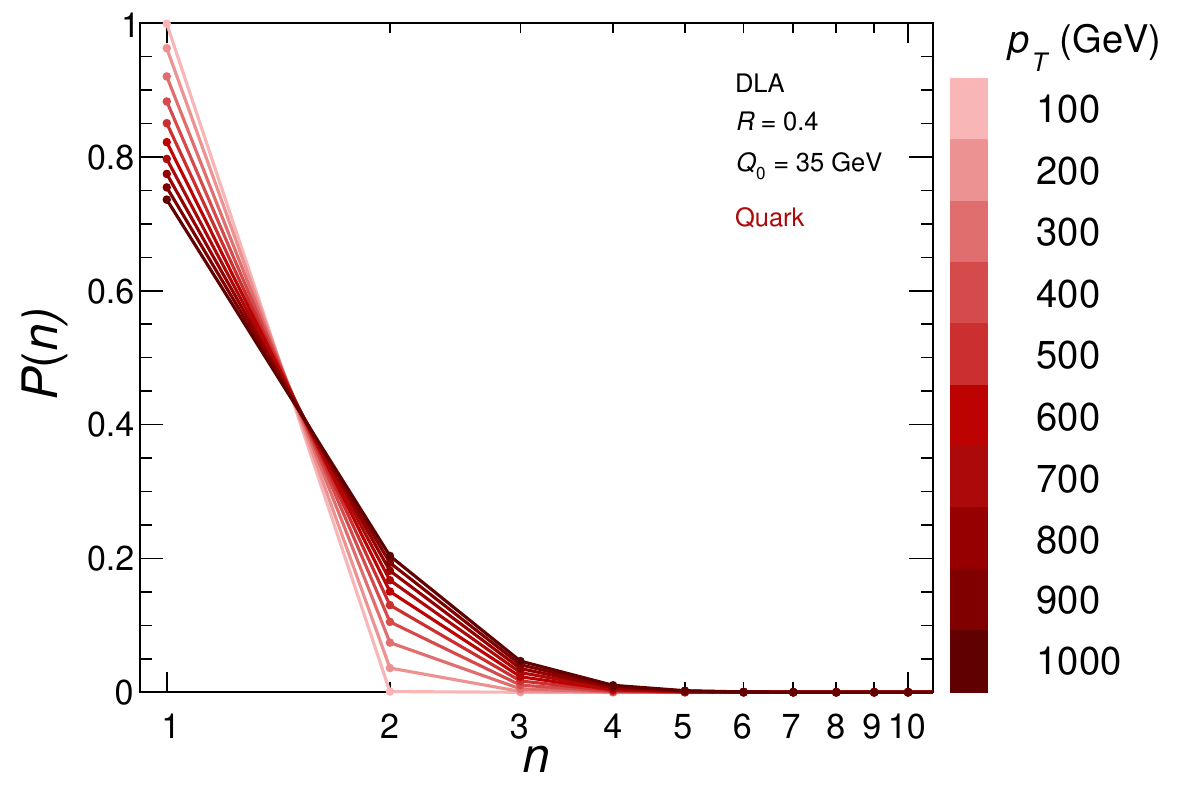}
    \includegraphics[width=0.49\textwidth]{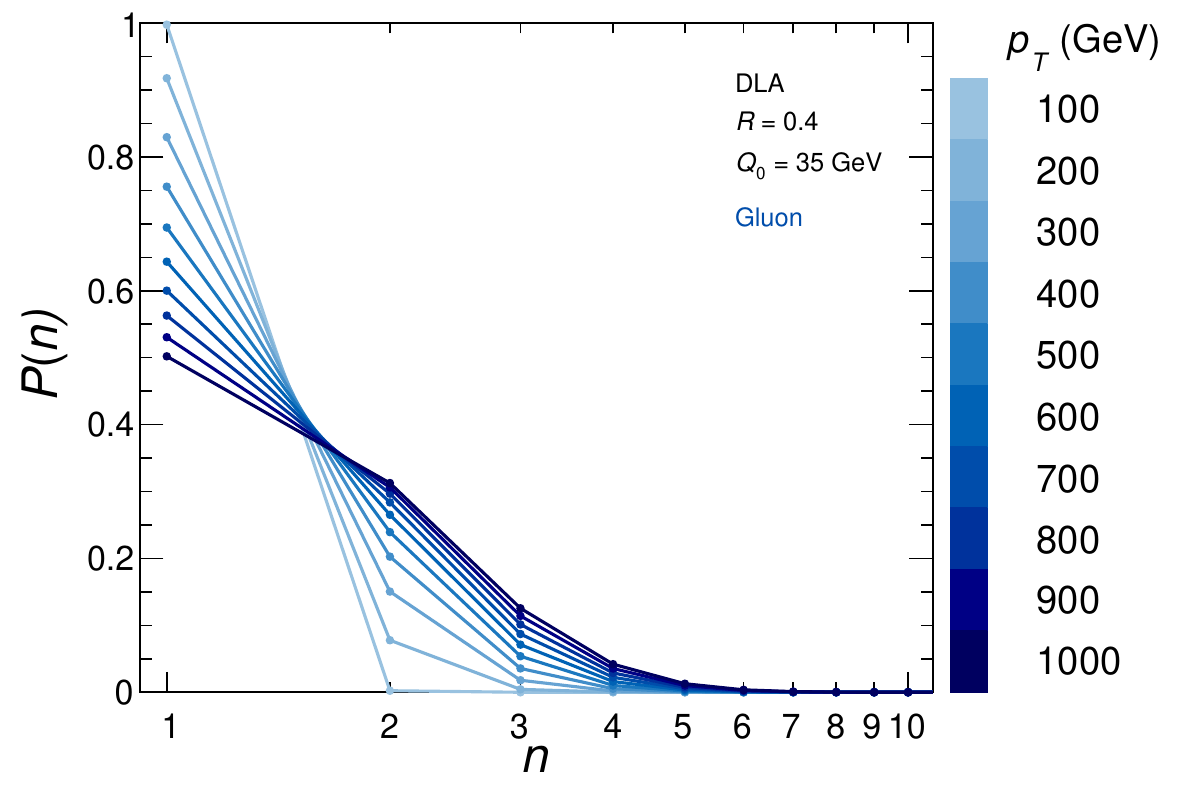}\\
    \includegraphics[width=0.49\textwidth]{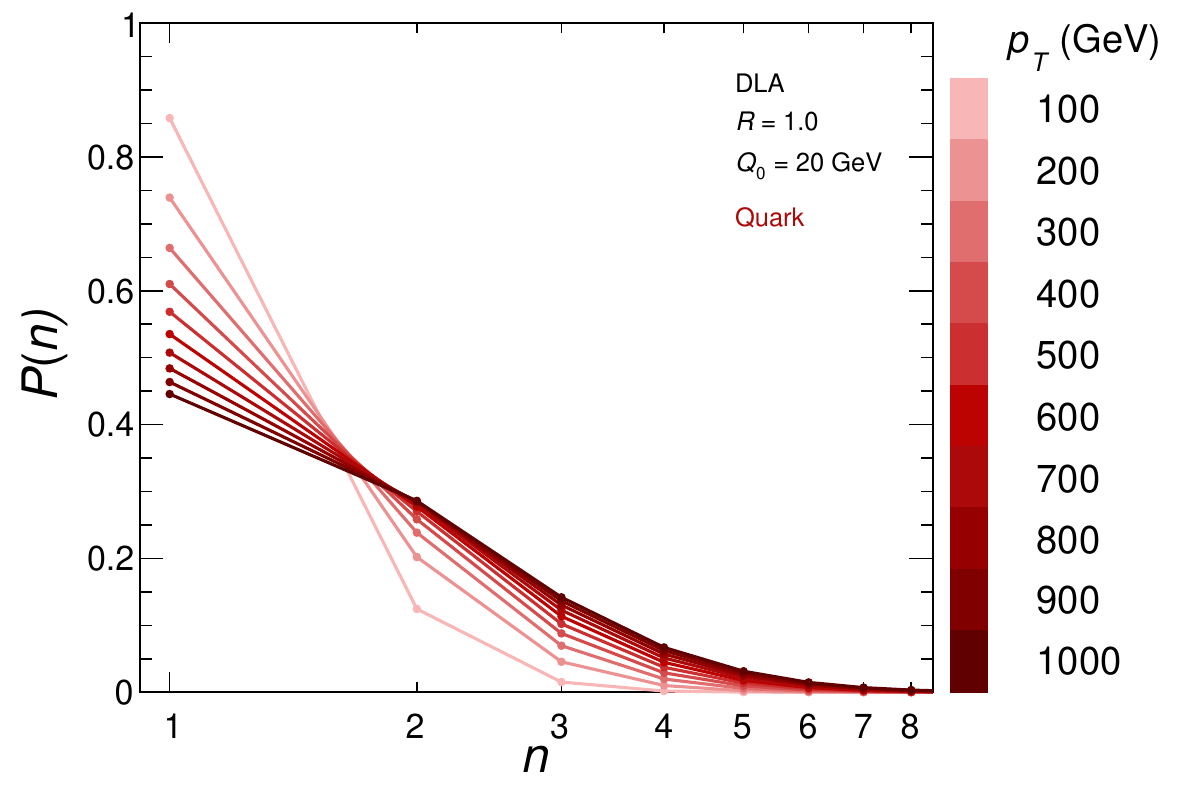}
    \includegraphics[width=0.49\textwidth]{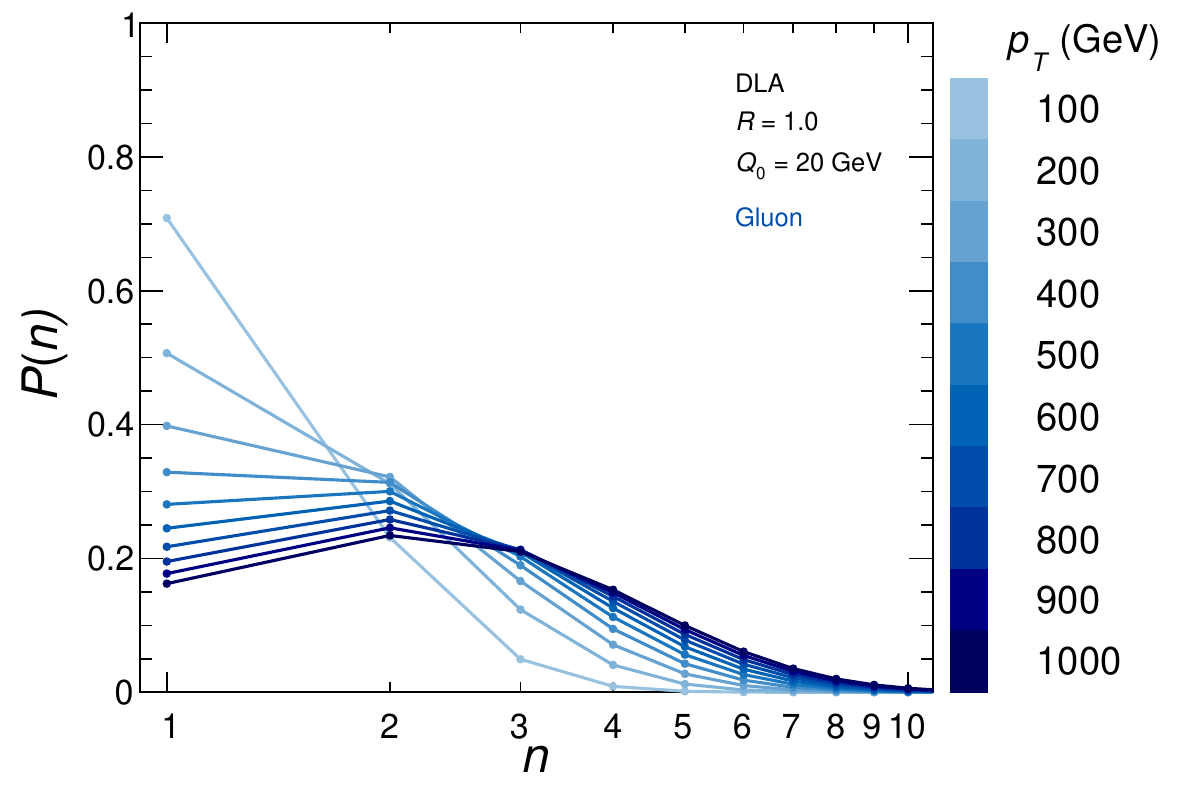}\\
    \caption{Parton multiplicity distributions for quark (red) and gluon (blue) jets. Top: $Q_0 = 35~\rm{GeV}$ and $R = 1.0$. Middle: $Q_0 = 35~\rm{GeV}$ and $R = 0.4$. Bottom: $Q_0 = 20~\rm{GeV}$ and $R = 1.0$. The different curves in each plot correspond to different values of $p_T$, ranging from $100~\rm{GeV}$ to $1000~\rm{GeV}$.}
    \label{fig:Pn-Q-R}
\end{figure}

Given that the values of $Q_0$ differ between $pp$ and $AA$ collisions, we first investigate the parton multiplicity probability distributions for higher values of $Q_0$ and different jet radius $R$ to facilitate the interpretation of our results when comparing with LHC data below. Figure~\ref{fig:Pn-Q-R} displays the resulting distributions for jets with transverse momenta ranging from $100~\rm{GeV}$ to $1000~\rm{GeV}$. Results for quark jets are shown by red lines in the left panels. Corresponding results for gluon jets are shown by blue lines in the right panels. The top two panels present the results obtained with $Q_0 = 35~\rm{GeV}$ and a large jet radius $R = 1.0$. It is observed that jets with higher transverse momentum generate a larger number of partons. This behavior reflects the increased phase space available for parton splittings. In addition, gluon jets consistently exhibit higher multiplicities than quark jets. This feature arises from the larger color factor associated with gluon emissions. In contrast, the middle two panels show the multiplicity distributions for the same value of $Q_0 = 35~\rm{GeV}$ but with a smaller jet radius $R = 0.4$. Compared to the large-radius case, the overall multiplicities are significantly reduced. This reduction can be attributed to the limited angular phase space of the smaller jet cone. Consequently, fewer partons are included in jet. The bottom two panels illustrate the impact of varying the parameter $Q_0$. In this case, the calculations are performed with $Q_0 = 20~\rm{GeV}$ and $R = 1.0$. When compared with the top panels, a clear enhancement of parton multiplicities is observed. This trend indicates that a smaller value of $Q_0$ increases the vacuum-like evolution stage. As a result, more partons are produced before medium-induced energy loss becomes dominant.

In summary, these results demonstrate that parton multiplicities are strongly influenced by both the jet cone size $R$ and $Q_0$. Smaller values of $Q_0$ and larger $R$ lead to higher multiplicities. This behavior highlights the importance of vacuum-like emissions in shaping the internal structure of the jet before significantly altering the medium.

\subsection{Mean medium-induced energy loss}
\label{sec:meanEloss}

Next, let us compare the energy loss of a single hard parton (that is, treating the jet as a single color charge) with the energy loss of multiple hard partons due to color decoherence. To be more specific, we study the mean medium-induced energy loss of jets for both cases, corresponding to eqs.~\eqref{eq:sigmaAA} and \eqref{eq:sigmaAA_de}, respectively.

In terms of the quenching weight distribution $D_i(\epsilon)$ in eq.~\eqref{eq:D}, the mean medium-induced energy loss for jets as a single color charge can be expressed as
\begin{align}\label{eq:dE0}
    \Delta E_i &= \int d\epsilon \, \epsilon \, D_i(\epsilon)\notag\\
    &=\sum_{n=0}^{\infty} \frac{1}{n!}
      \left[\prod_{k=1}^{n} \int d\omega_k \frac{dI_i(\omega_k)}{d\omega} \right]
      \bigg(\sum_{k=1}^{n} \omega_k\bigg)
      \exp\!\left[- \int_0^\infty d\omega\, \frac{dI_i(\omega)}{d\omega} \right] \notag\\
    &=\int d\omega \omega \frac{dI_i(\omega)}{d\omega} = \frac{\alpha_s N_c}{4}\hat{q}_i L^2 = \frac{\alpha_s C_i}{2}\omega_c.
\end{align}
Here, the index $i=q$ or $\bar{q}$ corresponds to quark jets, while $i=g$ refers to gluon jets. In this case, the impact of the jet substructure on energy loss cannot be identified.

On the other hand, when color decoherence effects are taken into account, this simplified picture no longer holds. The QCD medium is assumed to resolve partons with virtuality down to $Q_0$ during the shower evolution, while further splittings of these partons remain unresolved due to color coherence. As a result, the mean medium-induced energy loss becomes sensitive to the parton multiplicity inside the jet. In this case, the total mean energy loss receives contributions weighted by the probability of producing a given number of partons,
\begin{align}\label{eq:dE1}
    \Delta E_i
    &=P_i(1,p_TR) \int d\epsilon_1 \, \epsilon_1 D_i(\epsilon_1)\, \nonumber\\
    &\quad +\, \sum_{n=2}^{N} P_i(n,p_TR) \int d\epsilon_1 \, D_i(\epsilon_1) \left(\prod_{m=2}^{n} \int d\epsilon_m D_g(\epsilon_m)\right) \left(\sum_{k=1}^{n} \epsilon_k\right).
\end{align}
Here, the parton multiplicity probability for quark or gluon jets is evaluated numerically according to eq.~\eqref{eq:Pqg}, as in refs.~\cite{Duan:2025ngi, Duan:2025lvi}. In the numerical calculations of eq.~\eqref{eq:dE1}, we set $N = 15$. This choice ensures that the multiplicity probability approximately satisfies the normalization condition $\sum_{n=1}^{N} P_i(n, p_T R) = 1$, while remaining numerically tractable. Overall, contributions from higher multiplicities are numerically negligible in this setting.

\begin{figure}[htbp]
    \centering
    \includegraphics[width=0.49\textwidth]{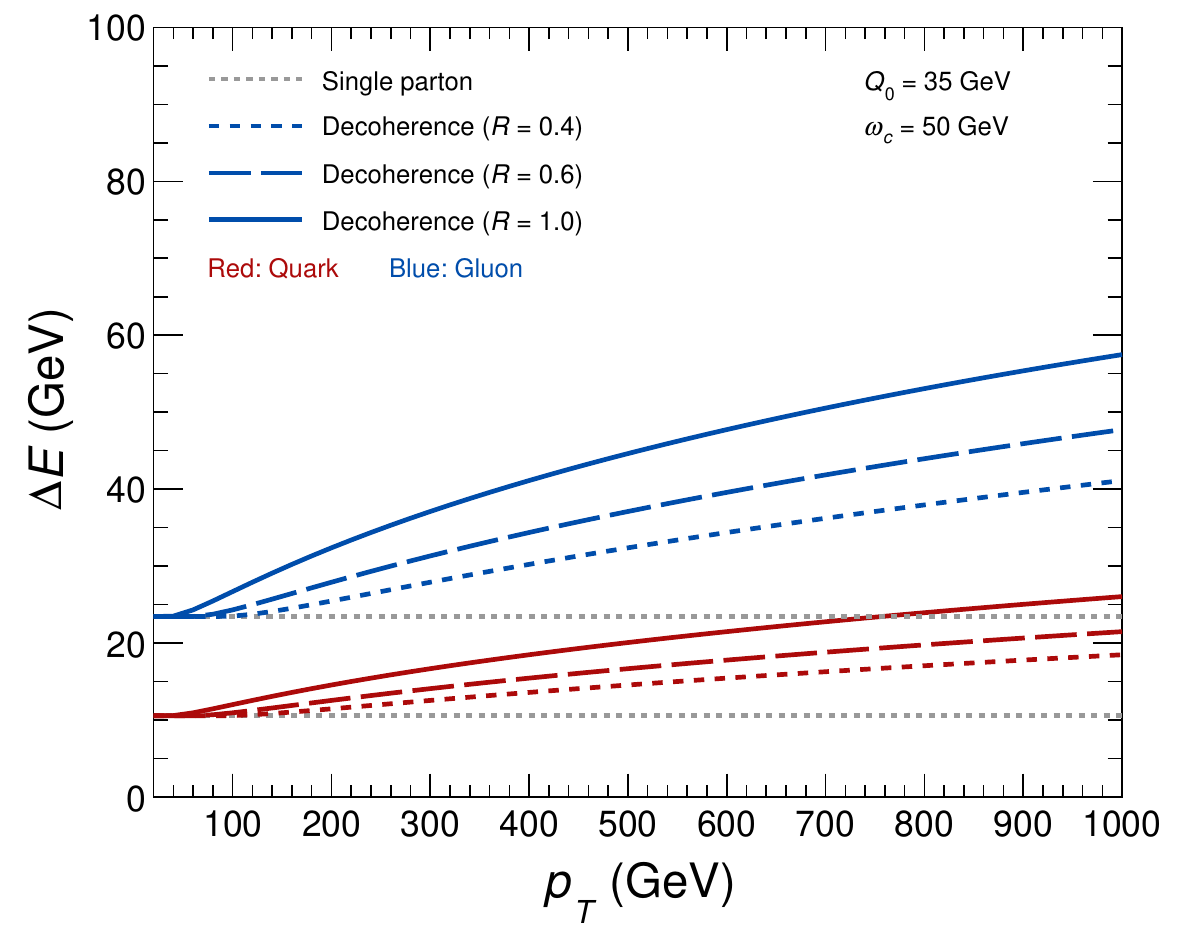}
    \includegraphics[width=0.49\textwidth]{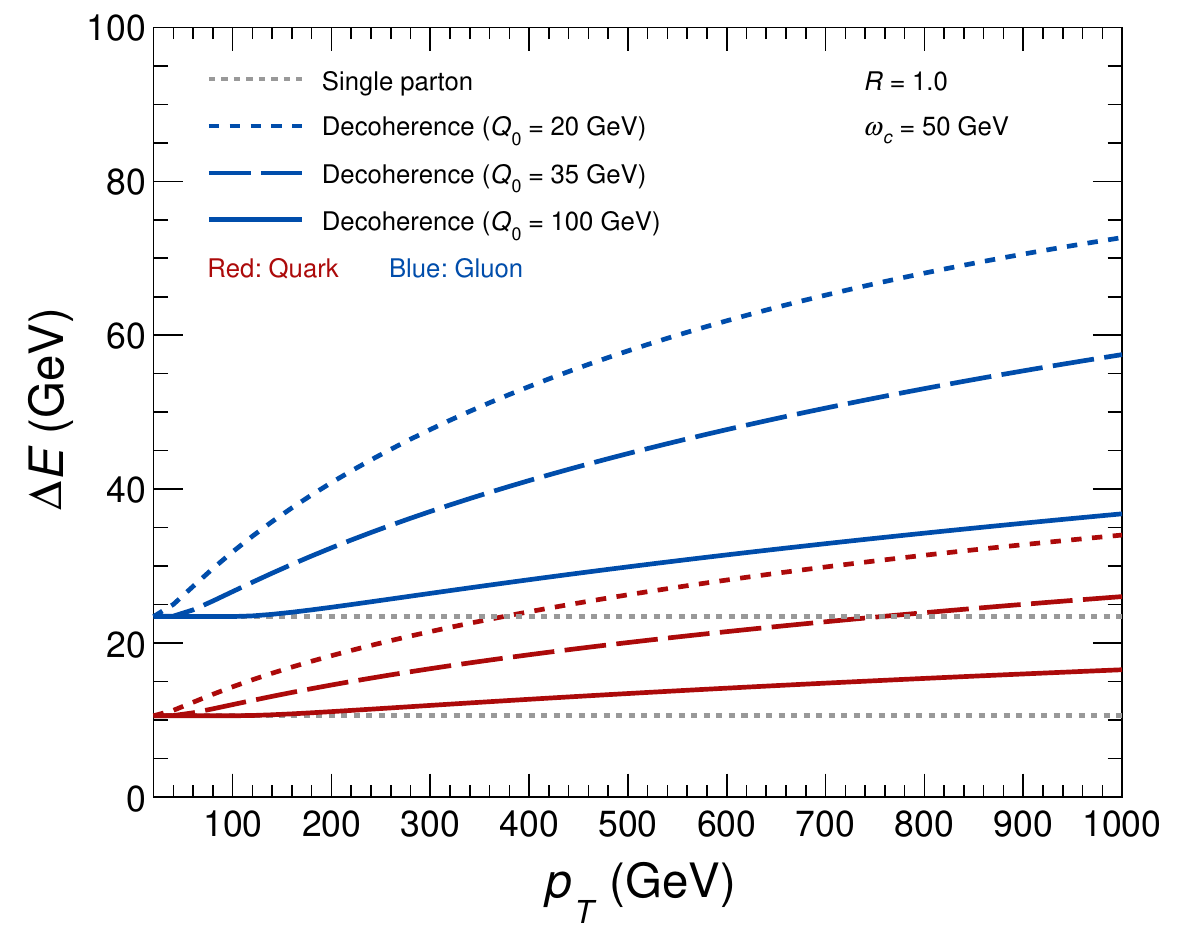}
    \caption{Mean medium-induced energy loss for $\omega_c = 50~\rm{GeV}$. Left: the dependence on the jet radius for $R = 0.4$, $0.6$, and $1.0$ with a fixed $Q_0 = 35~\rm{GeV}$. Right: the dependence on the color decoherence for $Q_0 = 20$, $35$, and $100~\rm{GeV}$ with a fixed jet radius $R = 1.0$.}
    \label{fig:eloss}
\end{figure}

We now investigate the dependence of the mean medium-induced energy loss on the parameters $R$ and $Q_0$. Figure~\ref{fig:eloss} shows the resulting mean medium-induced energy loss for a fixed characteristic gluon frequency $\omega_c = 50~\rm{GeV}$. In the figure, the gray dashed lines represent the results for a single parton obtained from eq.~\eqref{eq:dE0}. These curves are independent of the jet transverse momentum and therefore serve as a reference baseline. We confirm that the mean medium-induced energy loss between gluon and quark jets follows the ratio of $C_A/C_F = 9/4$, as expected from eq.~\eqref{eq:dE1}. This result follows directly from eqs.~\eqref{eq:qhat} and \eqref{eq:omegac}, where the jet quenching parameter $\hat{q}_i$ is proportional to $C_i$.

In contrast, the situation becomes much more complex for the color decoherence effect with multiple partons within jets. In addition to energy loss associated with the leading parton in each jet, which is of the same type as the parton initiated the jet, the jet also suffers energy loss due to medium-induced radiation of its gluon constitutents. In figure~\ref{fig:eloss}, the results of color decoherence are presented by the red lines for quark jets and the blue lines for gluon jets. Once color decoherence is included, a pronounced jet $p_T$ dependence emerges. In particular, the energy loss increases with increasing jet transverse momentum. This behavior originates from the larger parton multiplicities produced at higher $p_T$, which enter explicitly through the multiplicity distribution in eq.~\eqref{eq:dE1}. Furthermore, gluon jets experience larger energy loss than quark jets. This difference reflects the larger color factor associated with the radiation of gluons.

In addition, the left panel of figure~\ref{fig:eloss} illustrates the dependence on the jet radius for a fixed $Q_0 = 35~\rm{GeV}$. Calculations are performed for $R = 0.4$, $0.6$, and $1.0$. We observe that the mean energy loss increases monotonically with increasing jet radius. In particular, the largest value $R = 1.0$ leads to the strongest suppression. This trend can be understood as a consequence of the larger phase space available for vacuum-like emissions within wider jets. The right panel of figure~\ref{fig:eloss} presents the dependence on $Q_0$ for a fixed jet radius $R = 1.0$. Results are shown for $Q_0 = 20$, $35$, and $100~\rm{GeV}$. A clear reduction of the energy loss is observed as $Q_0$ increases. Smaller values of $Q_0$ allow the vacuum-like emissions to evolve over a longer virtuality range. Hence, more partons are produced before medium-induced radiation becomes dominant. Consequently, jets with smaller $Q_0$ experience larger energy loss.

Taken together, these results demonstrate that both the larger jet radius $R$ and smaller cutoff scale $Q_0$ enhance the mean medium-induced energy loss. This behavior is fully consistent with the multiplicity trends observed in figure~\ref{fig:Pn-Q-R}. It further highlights the central role of parton multiplicity as a consequence of color decoherence in determining the energy loss pattern of jets propagating through the QGP.

\section{Suppression of jets reconstructed from subjets with different cone sizes}
\label{sec4}

The cone-size dependence of jet suppression provides a sensitive probe for studying the microscopic mechanisms controlling jet-medium interactions~\cite{Mehtar-Tani:2017web, Mehtar-Tani:2021fud,CMS:2021vui,ATLAS:2023hso,Mehtar-Tani:2024jtd,Han:2025ukx}. In particular, large-radius jets with larger phase spaces contain more information about jet substructure and energy loss. In this section, we first determine the values of $Q_0$ and the jet quenching parameter at the initial time of the hydrodynamic evolution by fitting to the experimental measurements of the jet nuclear modification factor $R_{AA}$ for large-radius jets, as reported by ATLAS for 0–10\% PbPb collisions at $\sqrt{s_{NN}} = 5.02~\rm{TeV}$~\cite{ATLAS:2023hso}. We then provide quantitative predictions for the cone-size dependence of the suppression of jets with different cone sizes, reclustered from $R=0.2$ jets.

\subsection{Jet quenching parameter and $Q_0$}

In the previous section, we use the fixed $\omega_c$ to calculate mean medium-induced energy loss for illustration purpose. It is well known that for realistic heavy-ion collisions, the evolution of the medium can be well described by relativistic hydrodynamics. In the following phenomenological studies, we calculate $\omega_c$ by using the OSU (2+1)-dimensional viscous hydrodynamic model~\cite{Song:2007ux,Song:2008si,Song:2010mg}, which provides the local temperature $T(x)$ experienced by the propagating jet. In detail, the characteristic gluon frequency along the jet path is computed as~\cite{Salgado:2003gb}
\begin{align}\label{eq:omegac1}
    \omega_{c}(\vec{r},\phi)
    = \int d\tau\, \hat{q}(\tau)\, \tau.
\end{align}
Here, $\tau$ is the proper time along the  jet trajectory and $\hat{q}$ denotes the local jet transport coefficient for gluons, parameterized according to the temperature-dependent scaling
\begin{align}\label{eq:qhat0}
    \hat{q}(\tau) 
    = \hat{q}_{0} \left(\frac{T(\tau)}{T_0}\right)^3,
\end{align}
where $T(\tau)$ is the local medium temperature, $T_0$ is the temperature at the center of the QGP at the initial proper time $\tau_0$, and $\hat{q}_0$ corresponds to the jet transport coefficient for gluons at the same location and proper time. We take $\tau_0=0.6~\rm{fm/c}$ as initial time of QGP evolution and $T_c = 0.165~\rm{TeV}$ as the critical temperature stopping jet energy loss.

The color decoherence scale in QCD medium could be determined by comparing the transverse momentum associated with vacuum-like angular ordering to that generated by medium-induced momentum broadening, as done in refs.~\cite{Mehtar-Tani:2011hma,Casalderrey-Solana:2011ule,Blaizot:2012fh,Apolinario:2014csa,Caucal:2018dla,Caucal:2020xad}. In the reminder of this work, we instead take a phenomenological approach and treat $Q_0$ as a fitting parameter to be determined by experimental data. In particular, we focus on jets clustered from subjets with smaller jet radius, as measured in ref.~\cite{ATLAS:2023hso}, and treat such subjets with $R=0.2$ as color coherent. In this way, we can focus on investigating whether such a coarse-grained picture can capture the main features of the experimental data, thereby motivating and leaving the detailed calculations incorporating jet definitions at different $R$ in a time-dependent hydrodynamic background to future studies.

To quantitatively constrain the two parameters, namely $Q_0$ and the jet transport coefficient $\hat{q}_0$, we carry out a systematic $\chi^2/\rm{d.o.f.}$ analysis of the nuclear modification factor $R_{AA}$ for $R = 0.2$ jets and for $R=1.0$ jets reclustered from the $R=0.2$ jets in 0-10\% PbPb collisions at $\sqrt{s_{NN}} = 5.02~\rm{TeV}$ with jet rapidity restricted to $|y| < 2.0$~\cite{ATLAS:2023hso}. The $\chi^2/\rm{d.o.f.}$ is defined as
\begin{align}
    \chi^2/\rm{d.o.f.}
    = \frac{1}{N_{\rm{exp}}}\sum_{n=1}^{N_{\rm{exp}}} \frac{\bigl(Y_{\rm{exp}, n} - Y_{\rm{th}, n}\bigr)^2}{\sigma_{\rm{stat}, n}^2 + \sigma_{\rm{sys}, n}^2},
\end{align}
where $N_{\rm{exp}}$ denotes the total number of data points included in the fit, $Y_{\rm{exp,n}}$ and $Y_{\rm{th,n}}$ represent the experimental measurement and theoretical results using eq.~\eqref{eq:sigmaAA_de} for the $n$th observable, respectively, and $\sigma_{\rm{stat,n}}$ and $\sigma_{\rm{sys,n}}$ correspond to the reported statistical and systematic uncertainties for the experimental data.

\begin{figure}[htbp]
    \centering
    \includegraphics[width=0.49\textwidth]{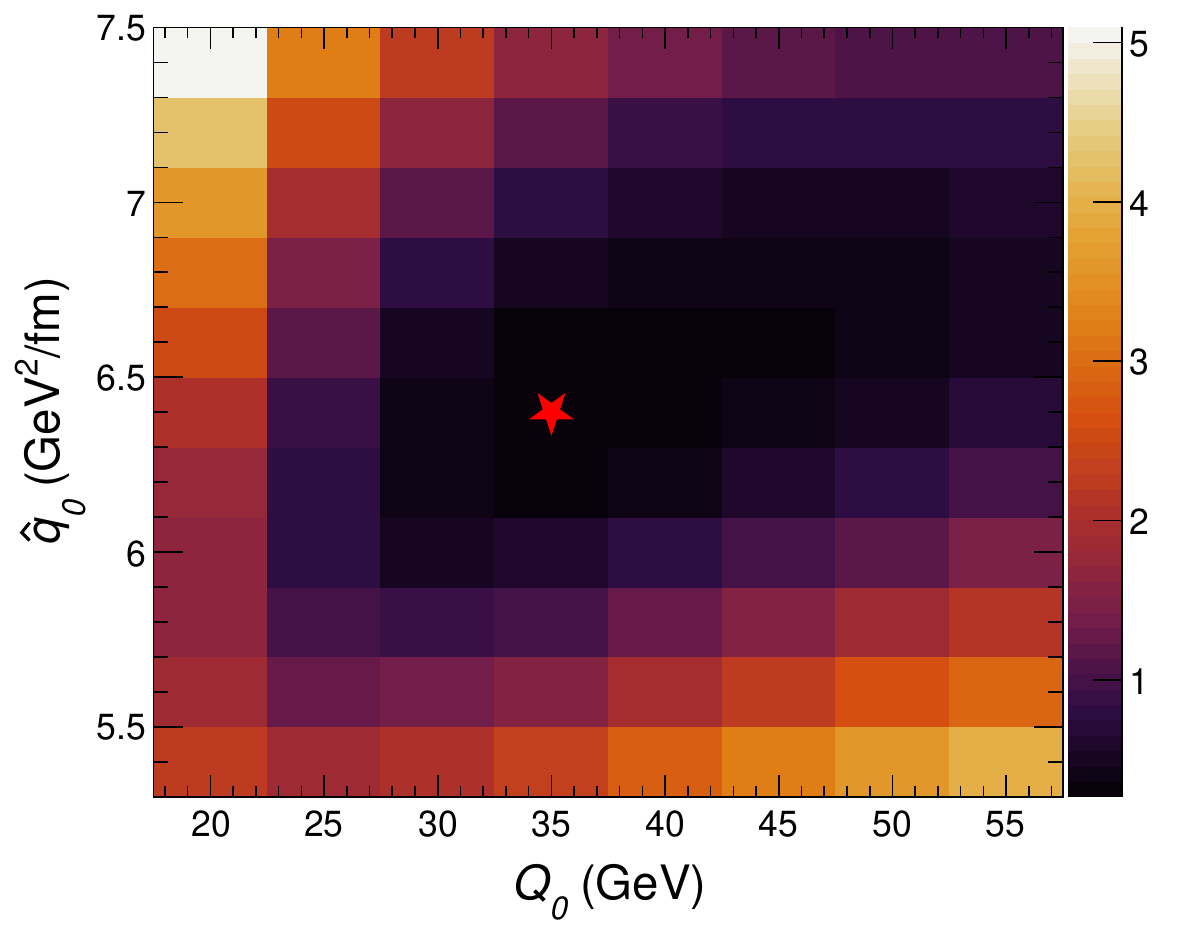}
    \includegraphics[width=0.49\textwidth]{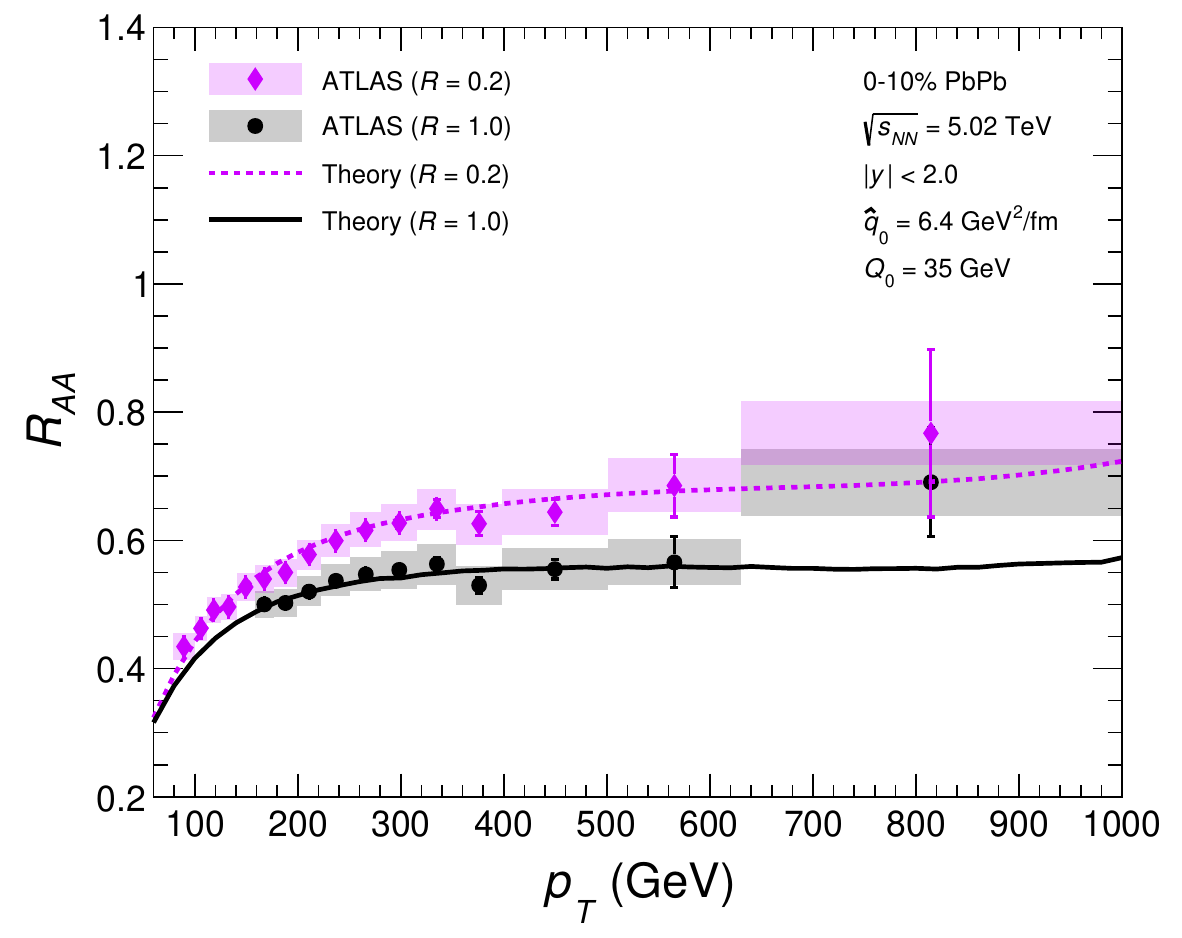}
    \caption{Nuclear modification factor $R_{AA}$ for $R=0.2$ jets and $R=1.0$ jets reclustered from $R=0.2$ jets in 0-10\% PbPb collisions at $\sqrt{s_{NN}} = 5.02~\rm{TeV}$. 
    Left: two-dimensional $\chi^2/\rm{d.o.f.}$ distribution obtained by scanning $Q_0$ and $\hat{q}_0$ using the experimental measurements of $R_{AA}$ reported in ref.~\cite{ATLAS:2023hso}. The red star denotes the minimum at $Q_0 = 35~\rm{GeV}$ and $\hat{q}_0 = 6.4~\rm{GeV^2/fm}$. 
    Right: theoretical results for $R_{AA}$ for these inclusive jets, compared with the ATLAS data~\cite{ATLAS:2023hso}.}
    \label{fig:RAAR10}
\end{figure}

The left panel of figure~\ref{fig:RAAR10} displays the resulting two-dimensional $\chi^2/\rm{d.o.f.}$ distribution obtained by scanning the parameter space in the ranges $Q_0 \in [20, 55]~\rm{GeV}$ and $\hat{q}_0 \in [5.4, 7.4]~\rm{GeV^2/fm}$ with ATLAS data~\cite{ATLAS:2023hso}. A minimum is observed at $Q_0 = 35~\rm{GeV}$ and $\hat{q}_0 = 6.4~\rm{GeV^2/fm}$ indicated by the red star marker in the figure. However, we note that a relatively broad region surrounding this minimum also yields comparably small $\chi^2/\rm{d.o.f.}$ values, implying a mild degree of parameter degeneracy within the considered domain. In particular, this correlation allows reasonable fits with smaller values of $\hat{q}_0$. Moreover, differences in the treatment of medium-induced radiation may contribute to the relatively larger value obtained in our analysis. Especially, the harmonic oscillator approximation used here can require larger values of $\hat{q}_0$ than approaches based on the full medium-induced radiation spectrum for a comparable amount of energy loss~\cite{Apolinario:2022vzg}. Nevertheless, the global minimum provides the optimal description of the ATLAS data within our present calculations. Hence, we adopt $Q_0 = 35~\rm{GeV}$ and $\hat{q}_0 = 6.4~\rm{GeV^2/fm}$ as the baseline parameter set. The right panel of figure~\ref{fig:RAAR10} presents our theoretical results of $R_{AA}$ for jet radii $R = 1.0$ and $R = 0.2$ in comparison with the ATLAS data~\cite{ATLAS:2023hso}. For these two jet radii, our theoretical results do show good agreement with the ATLAS data within the experimental uncertainties.

In the right panel of figure~\ref{fig:RAAR10}, a mild deviation between our theoretical results and experimental data is observed only in the highest $p_T$ bin for $R = 1.0$, where the experimental uncertainties are sizable. In our theoretical calculations, the $R_{AA}$ distributions exhibit an approximately flat behavior at high transverse momentum ($p_T > 300~\rm{GeV}$). However, a slight increase of $R_{AA}$ is observed for jets with $p_T > 850~\rm{GeV}$. This behavior originates from nuclear PDF modifications~\cite{Eskola:2021nhw} that are imprinted in the initial hard scattering processes in PbPb collisions, as discussed in previous studies~\cite{Pablos:2019ngg,Caucal:2020uic}. Large-radius jets with $R = 1.0$ are found to be more strongly suppressed than small-radius jets with $R = 0.2$, which leads to lower values of $R_{AA}$. In the ATLAS experiment, it is worth noting that large-radius jets ($R = 1.0$) are reconstructed by reclustering small-radius jets ($R = 0.2$) with $p_T > 35~\rm{GeV}$. This procedure effectively reduces the influence of the underlying event. On the other hand, it also prevents the recovery of quenched jet energy that is transported outside the $R = 0.2$ subjets. This effect is naturally described within our theoretical framework, in which hard partons are explicitly treated as jet substructure while medium-induced soft radiation is considered as energy loss regardless of whether it is inside or outside the large-radius jets. As a result, jets with larger radii naturally contain more subjets and experience larger medium-induced energy loss. Meanwhile, they make the growth of $R_{AA}$ very slow in the high jet $p_T$ region, avoiding the rapid approach of $R_{AA}$ to unity, as described above for the flat behavior.

\subsection{Cone-size dependence of jet suppression}

In this subsection, we explore further the cone-size dependence of $R_{AA}$ for jets reclustered from $R=0.2$ jets. This observable provides insight into how medium-induced radiation is distributed in angular phase space and how color decoherence affects energy loss inside the jet. This observable offers additional sensitivity to the interplay between vacuum-like emissions and medium-induced radiation and provides a direct test of this interplay.

\begin{figure}[htbp]
    \centering
    \includegraphics[width=0.49\textwidth]{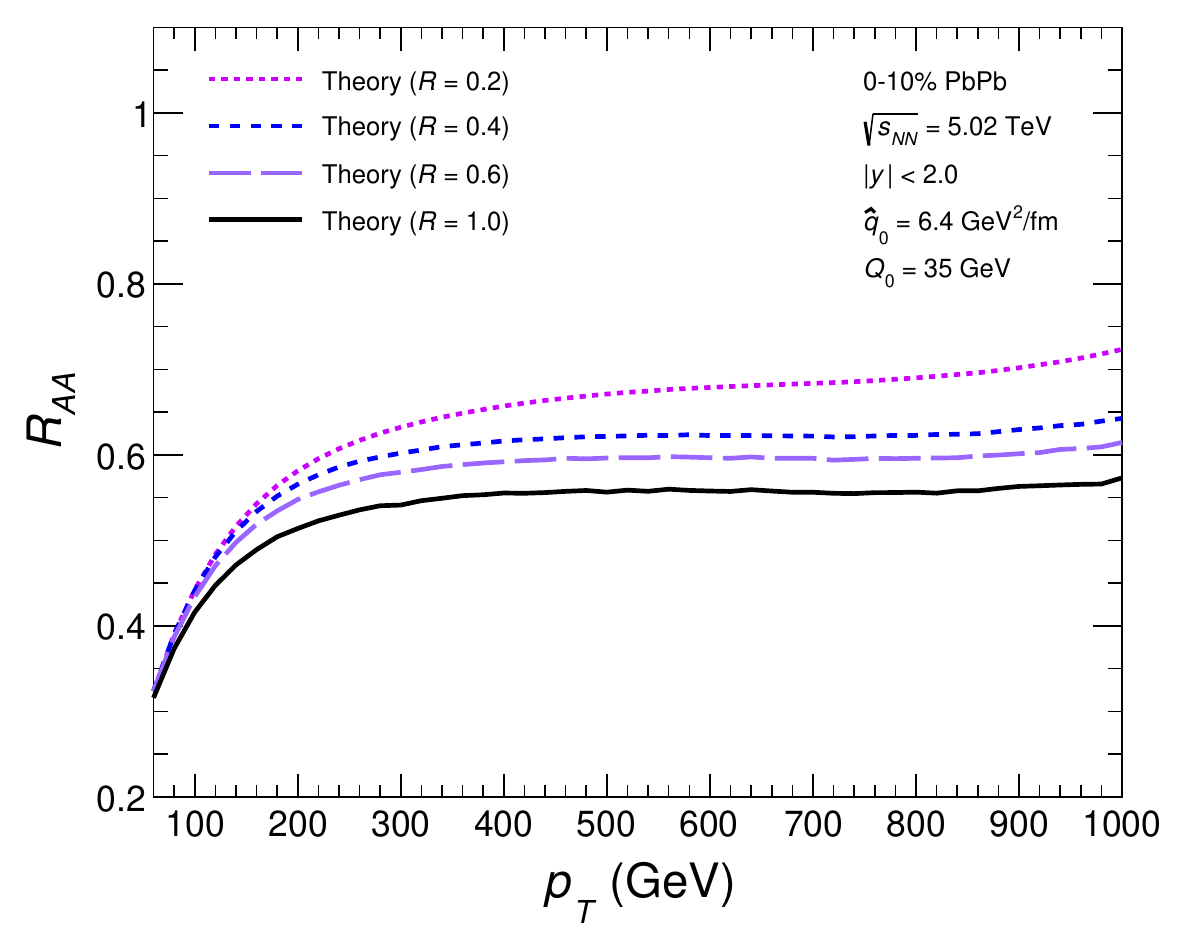}
    \includegraphics[width=0.49\textwidth]{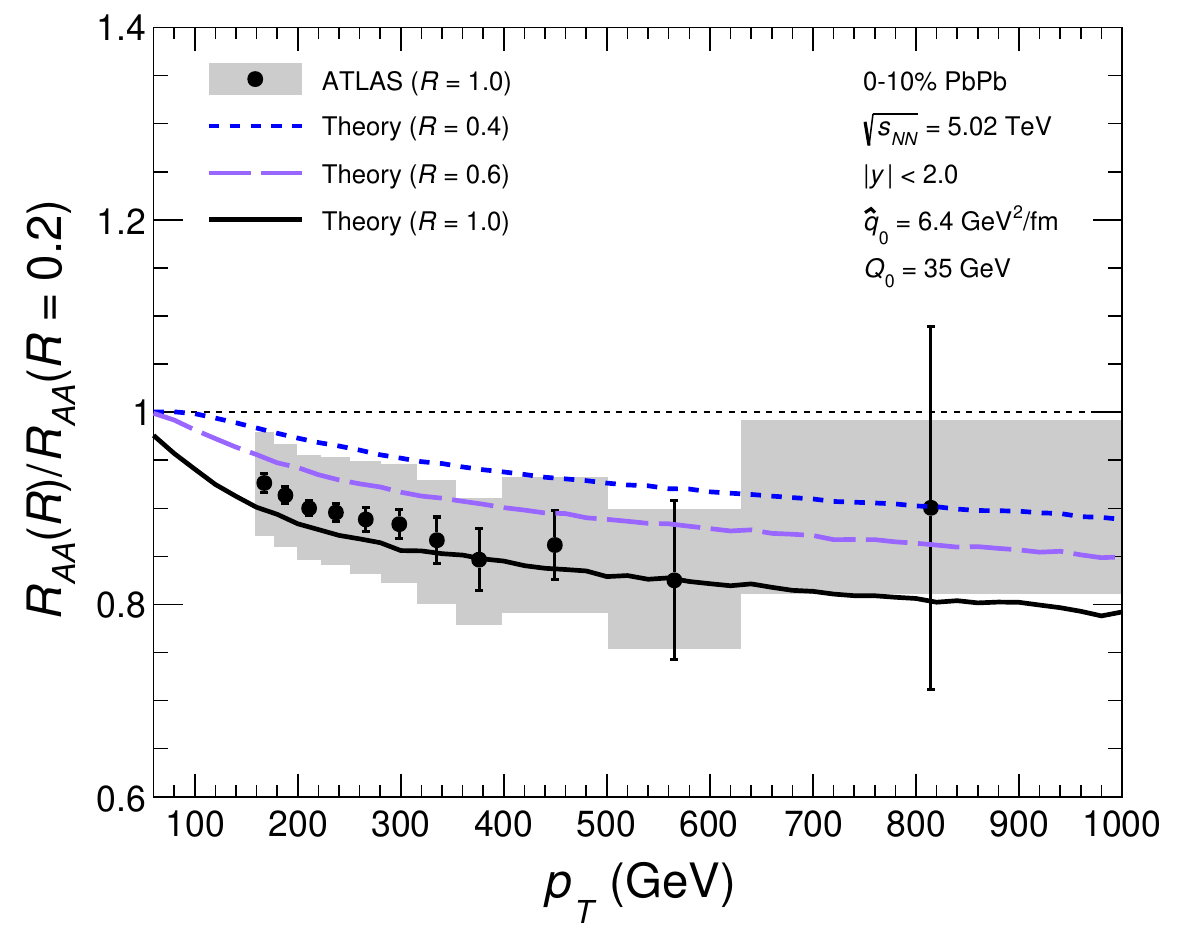}
    \caption{Cone-size dependence of the jet suppression $R_{AA}$ with jet substructure in 0-10\% PbPb collisions at $\sqrt{s_{NN}} = 5.02~\rm{TeV}$. Left: $R_{AA}$ for inclusive jets with radii $R = 0.4$, $0.6$, and $1.0$, reclustered from $R=0.2$ jets, in comparison with that for $R=0.2$. Right: the ratio of $R_{AA}$, shown in the left panel, normalized to the reference case $R = 0.2$. Here, $Q_0 = 35~\rm{GeV}$ and $\hat{q}_0 = 6.4~\rm{GeV^2/fm}$.}
    \label{fig:raa-r}
\end{figure}

We present predictions for additional cone-size dependence of such jet suppression using the baseline parameters $Q_0 = 35~\rm{GeV}$ and $\hat{q}_0 = 6.4~\rm{GeV^2/fm}$. The calculations are performed for the rapidity $|y| < 2.0$ in 0-10\% PbPb collisions at $\sqrt{s_{NN}} = 5.02~\rm{TeV}$. The left panel of figure~\ref{fig:raa-r} shows the nuclear modification factor $R_{AA}$ for inclusive jets with radii $R = 0.2$, $0.4$, $0.6$, and $1.0$, where the larger-radius jets are reclustered from $R=0.2$ jets. A clear ordering with respect to the jet radius can be observed: jets with larger cone sizes exhibit stronger suppression. This phenomenon suggests that jets with wider angular phase space include more partons that can interact with the medium and lose energy.

Furthermore, at low $p_T$, the $R_{AA}$ distributions obtained with color decoherence tend to converge toward a common value for all jet radii. This feature reflects the limited phase space for producing multiple resolved subjets in this kinematic region, where the probability of finding a single resolved subjet dominates and the energy loss becomes weakly dependent on the jet radius. This phenomenon has also been observed in low-$p_T$ jet measurements by the ALICE experiment~\cite{ALICE:2023waz}, where only mild dependence of $R_{AA}$ on $R$ is observed (note that the larger-radius jets in this measurement are not reclustered from smaller-radius jets). As the jet $p_T$ increases, the separation between different jet radii becomes more pronounced. This evolution indicates that color decoherence plays a more significant role and leading to more energy loss. This effect is more pronounced at larger jet radii due to the increased multiplicity of partons captured within the jet cone.

In addition, the right panel of figure~\ref{fig:raa-r} presents the ratio of $R_{AA}$ for different jet radii to the reference case $R = 0.2$. The lowest curve shows the ratio between $R = 1.0$ and $R = 0.2$. Our theoretical predictions reproduce the ATLAS measurements~\cite{ATLAS:2023hso} well over the entire $p_T$ range and exhibit the largest energy loss, as expected from our $\chi^2/\rm{d.o.f.}$ analysis above. This observation confirms that larger radius jets are more sensitive to medium-induced radiation. In comparison, the ratio between $R = 0.4$ and $R = 0.2$ is closer to unity, indicating smaller energy loss.

\section{Large-radius jet substructure suppression}
\label{sec5}

In this section, we investigate large-radius jet substructure suppression for the rapidity $|y| < 2.0$ in 0-10\% PbPb collisions at $\sqrt{s_{NN}} = 5.02~\rm{TeV}$. We focus on the multiplicity structure in parton shower due to vacuum-like emissions. Utilizing the theoretical framework established in the preceding sections, the total jet energy loss is decomposed into contributions from two different configurations: single subjet, treated as color coherent, and multiple subjets, treated as color decoherent. This separation provides a quantitative study to assess how color decoherence driven by parton multiplicity within the jet cone would alter the energy loss phenomenon and ultimately affects the observed nuclear modification factor $R_{AA}$.

\subsection{Mean medium-induced energy loss with substructure dependence}

For a parton produced at a hard scale $Q$, the number of partons down to the scale $Q_0$ within the parton shower initiated by the parton fluctuates, as described by $P_i(n, Q)$. This also implies that the parton initiating the jet can carry a virtuality smaller than $Q$, for example as represented by $P_i(1, Q)$. In this subsection, we isolate the energy loss for such jet events from that obtained by considering all $P_i(n, Q)$, as discussed in section~\ref{sec:meanEloss}. That is, we examine the mean medium-induced energy loss with this substructure dependence as a qualitative study before investigating the suppression of large-radius jet substructure.

We begin with the energy loss corresponding to a single subjet configuration where there is only one parton at scale $Q_0$, which acts as a single color charge propagating through the dense QCD medium due to color coherence. In this case, the contribution to the jet mean energy loss in eq.~\eqref{eq:dE1} from configurations containing a single parton $i$ can be written as
\begin{align}\label{eq:eloss1}
    \Delta E_{i,S}
    = P_i(1, p_T R) \int d\epsilon_1 \, \epsilon_1 D_i(\epsilon_1),
\end{align}
which corresponds to the first term in eq.~\eqref{eq:dE1}.

We now turn to the complementary case where, due to color decoherence, the jet effectively contains multiple resolved partons (subjets) from the perspective of the QCD medium. In this case, each parton loses energy independently according to individual color charges, and, accordingly, the contribution to the jet mean energy loss from multiple-subjet configurations is expressed as
\begin{align}\label{eq:elossN}
    \Delta E_{i,M}
    = \sum_{n=2}^{N} P_i(n, p_T R) \int d\epsilon_1 D_i(\epsilon_1) \left(\prod_{m=2}^{n} \int d\epsilon_m D_g(\epsilon_m)\right) \left(\sum_{k=1}^{n} \epsilon_k \right),
\end{align}
which corresponds to the second term in eq.~\eqref{eq:dE1}. Here, $n$ represents the total number of resolved partons, with the summation ranging from $n=2$ to the maximum value $N = 15$. For $n \ge 2$, $P_i(n, p_T R)$ gives the probability of $n$ partons. The first energy loss distribution $D_i(\epsilon_1)$ corresponds to the initial quark or gluon, while the subsequent $m$ distributions are taken as $D_g(\epsilon_m)$ for gluons. This choice reflects the dominance of gluon emission in parton shower within the DLA, where gluons lose more energy than quarks.

\begin{figure}[htbp]
    \centering
    \includegraphics[width=0.49\textwidth]{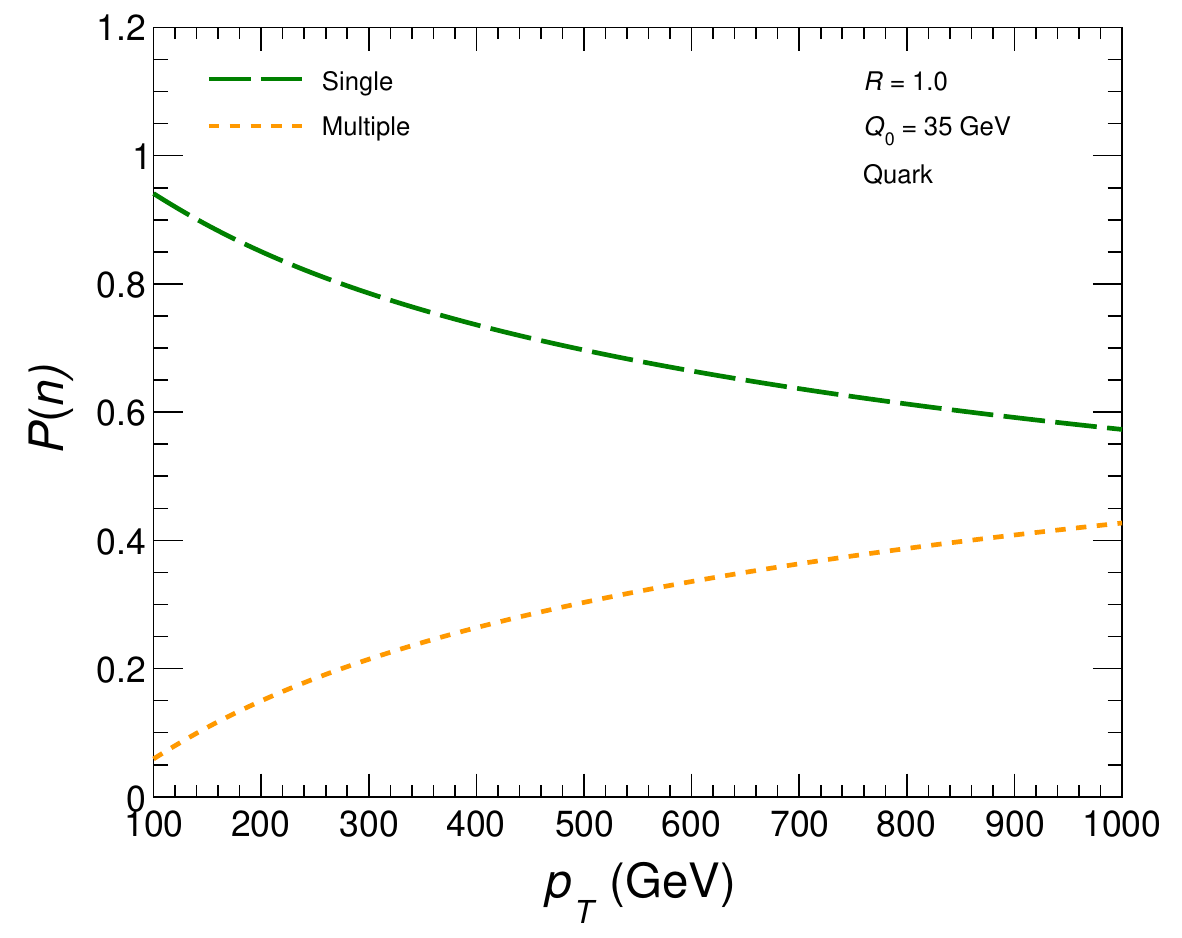}
    \includegraphics[width=0.49\textwidth]{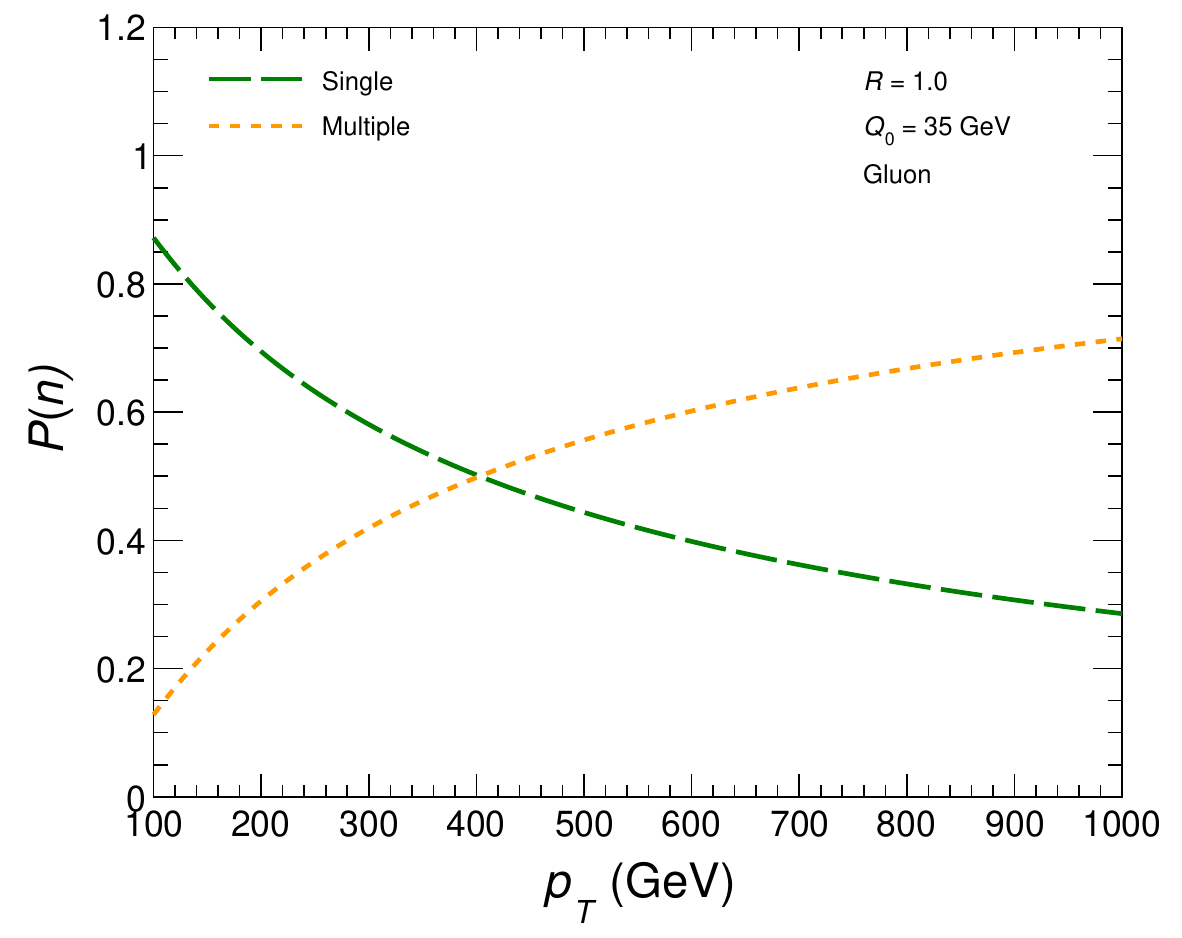}
    \caption{Probability for a jet containing single parton (green long-dashed lines) and multiple partons (orange dashed lines) as a function of jet $p_T$. Left: quark jets. Right: gluon jets.}
    \label{fig:Pn-sm}
\end{figure}

In this calculation, we employ the same parameter settings as in the previous section, where the large jet radius $R = 1.0$ and $Q_0 = 35~\rm{GeV}$. We begin by examining the multiplicity probability distributions $P_i(n, p_T R)$ as a function of jet $p_T$. Figure~\ref{fig:Pn-sm} displays these distributions, where the green long-dashed lines represents the probability for a jet containing a single resolved parton $P_i(1, p_T R)$ and the orange dashed lines denotes the probability for multiple resolved partons $\sum_{n=2}^N P_i(n, p_T R)$. The left panel shows the results for quark jets. Within the range of $100 < p_T < 1000~\rm{GeV}$, the probability for resolving a single parton consistently exceeds that for multiple partons. The single parton probability exhibits a slow, monotonic decline from approximately 94\% at $p_T = 100~\rm{GeV}$ to about 57\% at $p_T = 1000~\rm{GeV}$. Conversely, the right panel presents the results for gluon jets. A crossover point is observed near $p_T \approx 400~\rm{GeV}$, where the probabilities for resolving single parton and multiple partons becomes equal. At lower $p_T$ $\sim 100~\rm{GeV}$, the single parton dominates strongly at the 87\% level. However, this probability undergoes a rapid decrease with increasing $p_T$, dropping to roughly 29\% at $p_T = 1000~\rm{GeV}$, indicating a clear transition towards a significant multiple parton structure for high-$p_T$ gluon jets.

\begin{figure}[htbp]
    \centering
    \includegraphics[width=0.49\textwidth]{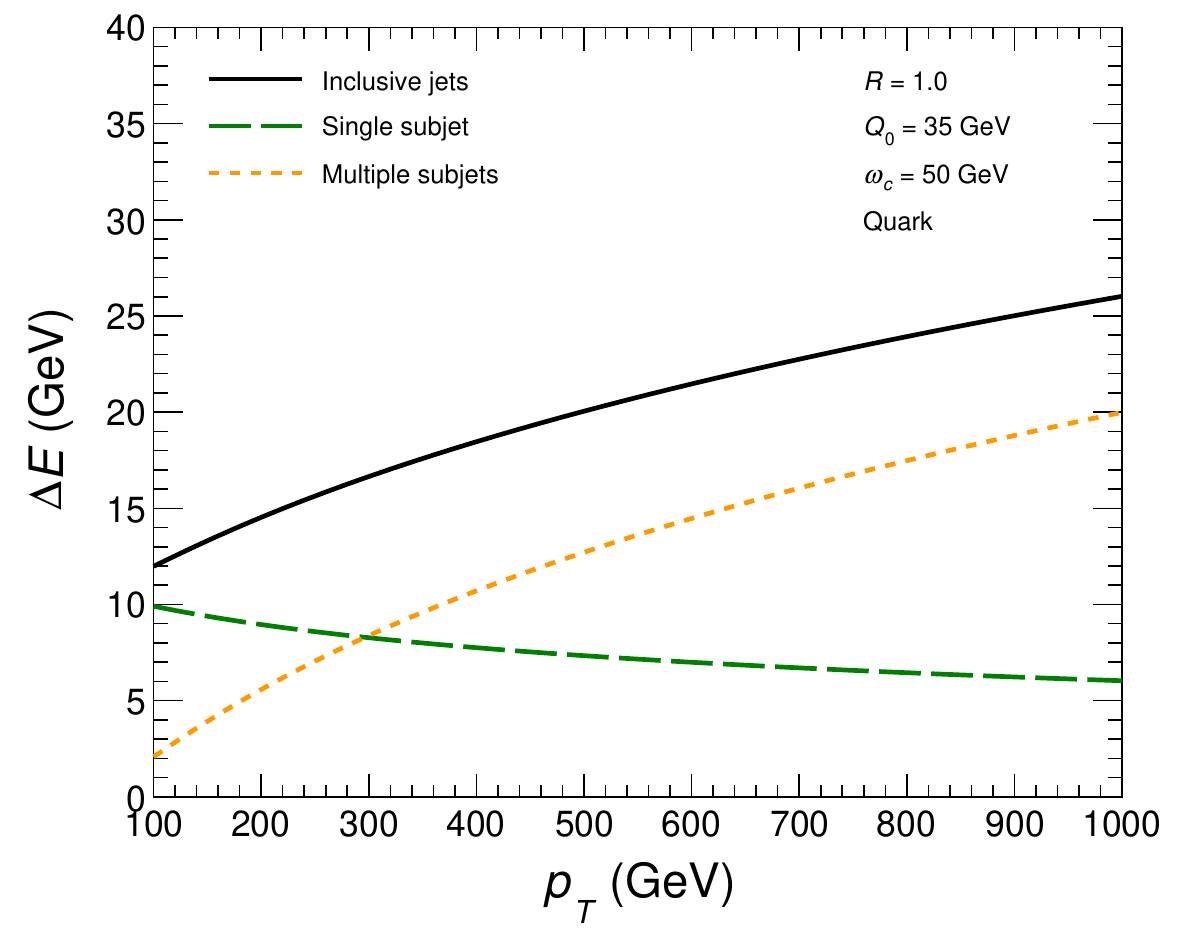}
    \includegraphics[width=0.49\textwidth]{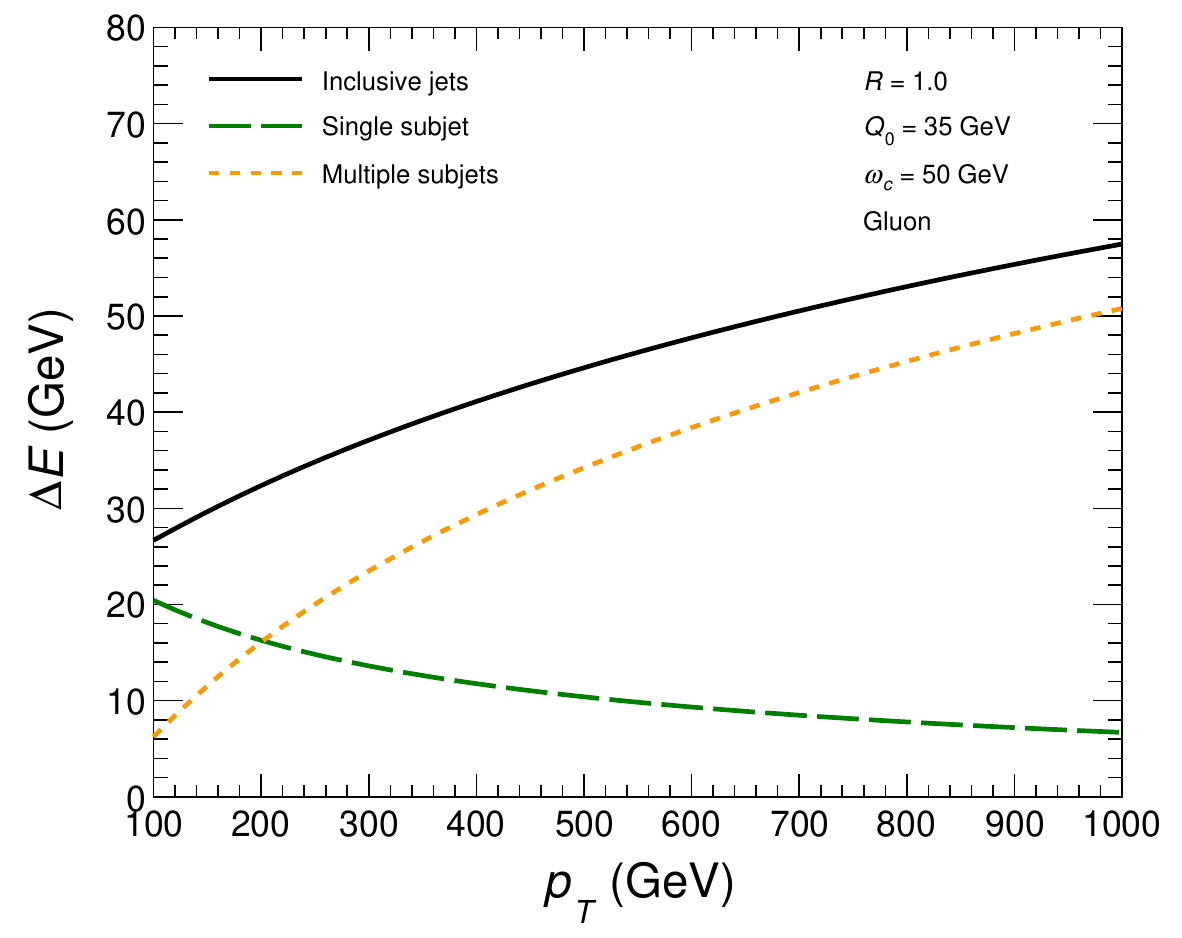}
    \caption{Mean medium-induced energy loss and its substructure dependence. The inclusive jet mean energy loss (black solid lines) are decomposed into contributions from single subjets (green long-dashed curves) and multiple subjets (orange dashed curves), as given in eqs.~\eqref{eq:eloss1} and \eqref{eq:elossN}, respectively. Left: quark jets. Right: gluon jets.}
    \label{fig:eloss-sm}
\end{figure}

Figure~\ref{fig:eloss-sm} presents the mean medium-induced energy loss, decomposed according to the number of subjets. Here, as in section~\ref{sec:meanEloss}, a fixed characteristic gluon frequency $\omega_c = 50~\rm{GeV}$ is used. In this figure, the black solid lines show the total energy loss for inclusive jets obtained from eq.~\eqref{eq:dE1}. This result is decomposed into the contribution from a single subjet (green long-dashed lines, from eq.~\eqref{eq:eloss1}) and that from multiple subjets (orange dashed lines, from eq.~\eqref{eq:elossN}).

The left panel of figure~\ref{fig:eloss-sm} shows the results of quark jets, which reveals a notable $p_T$-dependence in the loss mechanism. At low $p_T$ $\sim 100~\rm{GeV}$, the energy loss is overwhelmingly (about 83\%) dominatd by the single subjet configuration. This is due to the high probability of single subjet occurring at low $p_T$, as shown in figure~\ref{fig:Pn-sm}. In contrast, for multiple-subjet configurations, the mean energy loss increases significantly with $p_T$. This rise is related to the growing probability of multiple partons. In particular, the increase in the number of gluons can enhance significantly the overall energy loss. Consequently, at high $p_T$ $\sim 1000~\rm{GeV}$, multiple-subjet configurations contribute approximately 77\% of the total energy loss. The transition point where the two contributions are equal occurs near $p_T \approx 300~\rm{GeV}$ for quark jets.

The right panel of figure~\ref{fig:eloss-sm} displays the corresponding results for gluon jets. Although the qualitative trends are similar, shifting from single-subjet dominance at low $p_T$ to multiple-subjet dominance at high $p_T$, the quantitative differences are significant. The single subjet contribution constitutes only about 77\% of the total energy loss at $p_T = 100~\rm{GeV}$. This fraction is lower than that for quark jets due to the steeper decline of $P_g(1, p_T R)$. Correspondingly, the multiple-subjet contribution rises sharply, accounting for nearly 88\% of the energy loss at $p_T = 1000~\rm{GeV}$. The crossing point where the two contributions are equal is found at a lower value, $p_T \approx 200~\rm{GeV}$, for gluon jets. Compared to quark jets, this shift to lower $p_T$ highlights the greater energy loss effect of gluon jets, as shown in figure~\ref{fig:Pn-sm}. These results highlight that the jet substructure, determined by the flavor of the initial jet and the jet $p_T$, plays a crucial role in determining the total energy loss in the medium.

\subsection{Large-radius jet suppression with substructure dependence}

Building upon the energy loss formalism established in the preceding subsection, we now derive the $R_{AA}$ contribution arising from single and multiple subjet configurations of large-radius jets. This explicit decomposition allows for a more detailed investigation of how color coherence and decoherence manifest in the jet suppression.

For jets with a single subjet, the differential cross section in $AA$ collisions can be expressed as 
\begin{align}\label{eq:sigmaAA1}
    \frac{d\sigma^{AA}_S}{d^2 \vec{b}\, dp_T} &= \int d^2 \vec{r}\, T_A(\vec{r}+\vec{b}/2) T_B(\vec{r}-\vec{b}/2)\int \frac{d\phi}{2\pi} \nonumber\\
    &\quad \times \Bigg[\sum_{i} P_i(1,p_T^{\prime}R) \int d\epsilon_1 D_i(\epsilon_1) \left.\frac{d\sigma_i^{NN}}{dp_T^{\prime}}\right|_{p_T^{\prime}=p_T+\epsilon_1}\Bigg].
\end{align}
Serving as the baseline reference, the differential jet cross section for single subjet in pp collisions is given by
\begin{align}
   \frac{d\sigma_S^{pp}}{dp_T} = \sum\limits_{i} P_i(1,p_TR) \frac{d\sigma_i^{pp}}{dp_T}.
\end{align}
Accordingly, the corresponding nuclear modification factor follows directly as
\begin{align}
    R_{AA,S} =
    \frac{1}{T_{AB}(\vec{b})}
    \frac{d\sigma^{AA}_S/d^2\vec{b}\,dp_T}{d\sigma^{pp}_S/dp_T}.
\end{align}
This expression quantifies the suppression effect caused solely by the color coherence of the unresolved jet propagating in the medium.

In contrast, for jet events where the parton shower decoheres, the energy loss arises from multiple subjets. Each subjet loses energy independently according to its quenching weight. The differential cross section in $AA$ collisions must account for the multiplicity probability and the overall energy loss of all subjets. It can be expressed as
\begin{align}\label{eq:sigmaAAN}
    \frac{d\sigma^{AA}_M}{d^2 \vec{b}\, dp_T} &= \int d^2 \vec{r}\, T_A(\vec{r}+\vec{b}/2) T_B(\vec{r}-\vec{b}/2) \int\frac{d\phi}{2\pi}
    \notag\\
    &\quad \times \Bigg[\sum_{n=2}^{N} \sum_{i} P_i(n,p_T^{\prime}R)
    \int d\epsilon_1 D_i(\epsilon_1) \left(\prod_{m=2}^{n} \int d\epsilon_m D_g(\epsilon_m)\right) \left.\frac{d\sigma_i^{NN}}{dp_T^{\prime}}\right|_{p_T^{\prime}=p_T+\sum_{k=1}^{n} \epsilon_k}\Bigg].
\end{align}
Similarly, the differential jet cross section for multiple subjets in pp collisions is given by
\begin{align}
    \frac{d\sigma_M^{pp}}{dp_T} = \sum_{n=2}^{N}  \sum_{i} P_i(n,p_TR) \frac{d\sigma^{pp}_i}{dp_T}.
\end{align}
And the corresponding nuclear modification factor is given by
\begin{align}
    R_{AA,M} =
    \frac{1}{T_{AB}(\vec{b})}
    \frac{d\sigma^{AA}_M/d^2\vec{b}\,dp_T}{d\sigma^{pp}_M/dp_T}.
\end{align}
This component captures the suppression pattern for jets that effectively fragment into several independent color charges within the medium.

\begin{figure}[htbp]
    \centering
    \includegraphics[width=0.6\textwidth]{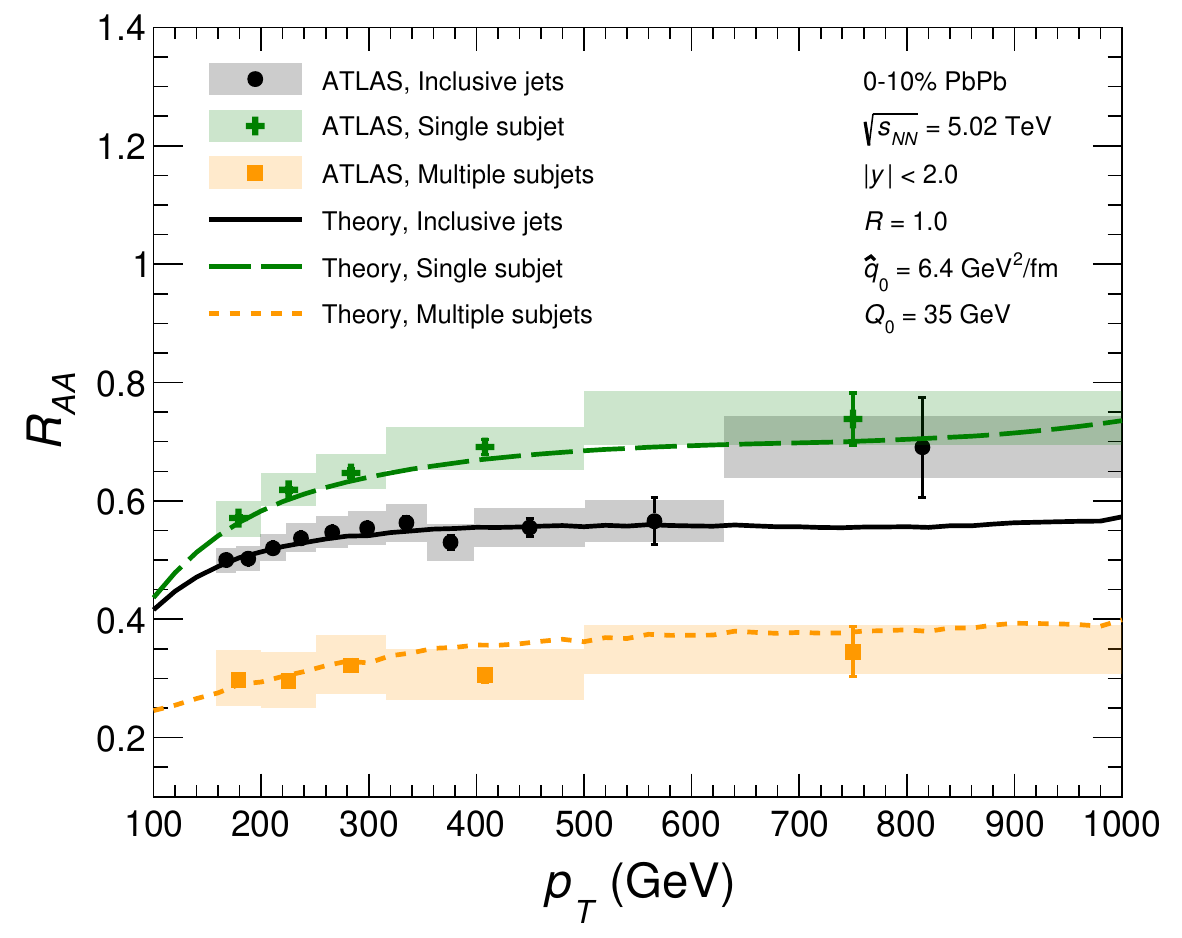}
    \caption{The nuclear modification factor $R_{AA}$ as a function of jet $p_T$ for jet substructure components with $R = 1.0$ and $|y| < 2.0$ in 0–10\% PbPb collisions at $\sqrt{s_{NN}} = 5.02~\rm{TeV}$. The calculation uses $Q_0 = 35~\rm{GeV}$ and $\hat{q}_0 = 6.4~\rm{GeV^2/fm}$. The black solid line shows the results for inclusive jets. The green long-dashed and orange dashed lines represent the contributions from single subjet and multiple subjets, respectively. The ATLAS data points~\cite{ATLAS:2023hso} shown in the corresponding colors are included for comparison.}
    \label{fig:RAA-SM}
\end{figure}

The $R_{AA}$ results for jets containing single subjet and multiple subjets are presented in figure~\ref{fig:RAA-SM}. We consider large-radius jets with $R = 1.0$ in the rapidity region $|y| < 2.0$ for 0–10\% PbPb collisions at $\sqrt{s_{NN}} = 5.02~\rm{TeV}$. The calculation employs the baseline parameters $Q_0 = 35~\rm{GeV}$ and $\hat{q}_0 = 6.4~\rm{GeV^2/fm}$. Several key features are obvious. First, our theoretical predictions for both single subjet and multiple subjets show good agreement with the ATLAS data within experimental uncertainties~\cite{ATLAS:2023hso}. More importantly, a significant separation exists between the $R_{AA}$ curves for single subjet (green long-dashed line) and multiple subjets (orange dashed line). Jets characterized by a single subjet exhibit the weakest suppression across the entire $p_T$ range. This result is consistent with the coherent propagation as a single color charge which experiences a relatively lower energy loss. Conversely, jets containing multiple resolved subjets show significantly stronger suppression. This enhanced suppression directly stems from the cumulative independent energy loss of each subjet. Furthermore, the $R_{AA}$ for all jet definitions displays an approximately flat $p_T$ dependence ($p_T > 300~\rm{GeV}$).

In addition, the separation provides an important signature of color decoherence in QGP. It is important to emphasize that, within our framework, this difference is not generated by soft radiation falling outside the $R=0.2$ subjet cones. Such radiation is not included in the reconstructed subjet energy in either our calculation or the experimental measurement. Rather, the separation between single-subjet and multiple-subjet contributions originates from the different number of medium-resolved subjets that experience independent energy loss. Although the present implementation does not explicitly track the angular distribution of medium-induced radiation inside the full jet radius $R=1.0$, the impact of this approximation is hence expected to be less significant for the subjet-dependent observables considered here than for inclusive jets with the same large radius, where the detailed angular redistribution of radiated energy plays a more direct role.

\section{Summary and outlook}
\label{sec6}

In this work, we have proposed a theoretical framework to describe the dynamical interplay between vacuum-like jet evolution and medium-induced interactions. Within this framework, the hard scattering processes are described using leading-order differential cross sections, vacuum-like emissions are evaluated in the double logarithmic approximation with running coupling through parton multiplicity probabilities, and medium-induced gluon radiations are formulated using the BDMPS-Z formalism. We interpret the infrared virtuality scale $Q_0$ in the vacuum-like jet evolution as an effective boundary between color-coherent and color-decoherent regimes in jet evolution within the medium.  By combining the vacuum-like shower evolution and medium-induced energy loss in this way, we have analyzed in detail the effects of color decoherence and virtuality on the jet nuclear modification factor $R_{AA}$.

For comparison, we first calculate medium-induced energy loss of a jet in QCD matter. We find that gluon jets experience a larger energy loss than quark jets, which arises from different color factors. We also find that the energy loss increases with increasing jet transverse momentum $p_T$, as higher-$p_T$ jets produce larger parton multiplicities during the vacuum-like emissions. Furthermore, we have examined the dependence of the energy loss on jet cone size $R$ and on $Q_0$. Our theoretical results demonstrate that both a larger jet radius and a smaller virtuality cutoff lead to stronger energy loss, since they contain more parton multiplicity.

Then, we employ the OSU (2+1)-dimensional viscous hydrodynamic model to describe the QGP evolution. Within this framework, we investigate $R_{AA}$ in 0-10\% PbPb collisions at $\sqrt{s_{NN}} = 5.02~\rm{TeV}$, focusing on its dependence on the jet cone size. Our results are compared with the ATLAS data~\cite{ATLAS:2023hso}. Using the parameter set $Q_0 = 35~\rm{GeV}$ and $\hat{q}_0 = 6.4~\rm{GeV^2/fm}$ with $\hat{q}_0$ the jet quenching parameter for gluons at initial proper time $\tau_0=0.6~\rm{fm}$, our theoretical results achieve good agreement with the ATLAS measurements for jet radii $R = 1.0$ and $R = 0.2$. We find that wider jets are found to be more strongly suppressed than narrow jets because they contain more subjets resolved by the medium. In this context, the suppression of large-radius jet substructure provides a sensitive probe of color decoherence effects in the medium.

We further decompose the total jet energy loss into contributions from a single subjet with color coherence and from multiple subjets with color decoherence. We find that our theoretical predictions for $R_{AA}$ containing either single subjet or multiple subjets are in good agreement with the ATLAS measurements~\cite{ATLAS:2023hso}. Meanwhile, the $R_{AA}$ result associated with a single subjet exhibits the weakest suppression across the entire $p_T$ range, whereas the $R_{AA}$ results for multiple resolved subjets shows a significantly stronger suppression. This clear separation between the two contributions provides a crucial signature of color decoherence in QGP.

Finally, we emphasize that our phenomenological approach, which incorporates vacuum-like emissions, BDMPS-Z energy loss, and a realistic QCD medium described by hydrodynamics, provides promising results. Nonetheless, there is clearly more to be done. It remains to be seen whether this framework can consistently describe additional jet observables, especially including substructure measurements. In particular, extending the formalism to retain the angular correlations between resolved subjets would allow a more differential study of jet substructure observables, such as subjet opening-angle distributions~\cite{ATLAS:2023hso}. Ultimately, it would be desirable to include the interplay between vacuum-like and medium-induced radiation within a unified framework derived from QCD first principles, accounting for the dependence of jet radius and reclustering procedures. We leave these interesting and challenging questions for future studies.

\begin{acknowledgments}
We thank Prof. Chun Shen for helpful discussions on the hydrodynamic model, and Prof. Daniel Pablos and Dr. Arjun Kudinoor for providing support with the ATLAS data.
X.D. thanks Prof. Weiyao Ke for helpful discussions.
This work is supported by the European Research Council under project ERC-2018-ADG-835105 YoctoLHC; by Maria de Maeztu excellence unit grant CEX2023-001318-M and project PID2023-152762NB-I00 funded by MICIU/AEI/10.13039/501100011033; and by ERDF/EU. It has received funding from Xunta de Galicia (CIGUS Network of Research Centres). 
X.D. and G.M. are supported by the National Natural Science Foundation of China under Grants No.12147101, No. 12547102, No. 12325507, the National Key Research and Development Program of China under Grant No. 2022YFA1604900, and the Guangdong Major Project of Basic and Applied Basic Research under Grant No. 2020B0301030008. 
L.C. is supported by the Marie Sk\l{}odowska-Curie Actions Postdoctoral Fellowships under Grant No.~101210595. 
B.W. acknowledges the support of the Ram\'{o}n y Cajal program with the Grant No. RYC2021-032271-I and the support of Xunta de Galicia under the ED431F 2023/10 project.
\end{acknowledgments}

\bibliographystyle{apsrev4-2}
\bibliography{ref.bib}

\end{document}